\documentclass[preprint,journal]{vgtc}       % preprint (journal style)

\ifpdf%                                % if we use pdflatex
  \pdfoutput=1\relax                   % create PDFs from pdfLaTeX
  \usepackage{graphicx}                % allow us to embed graphics files
  \DeclareGraphicsExtensions{.pdf,.png,.jpg,.jpeg} % for pdflatex we expect .pdf, .png, or .jpg files
\else%                                 % else we use pure latex
  \usepackage{graphicx}                % allow us to embed graphics files
  \DeclareGraphicsExtensions{.eps}     % for pure latex we expect eps files
\fi%

\graphicspath{{figures/}{pictures/}{images/}{./}} % where to search for the images

\usepackage{microtype}                 % use micro-typography (slightly more compact, better to read)
\PassOptionsToPackage{warn}{textcomp}  % to address font issues with \textrightarrow
\usepackage{textcomp}                  % use better special symbols
\usepackage{mathptmx}                  % use matching math font
\usepackage{times}                     % we use Times as the main font
\usepackage{cite}                      % needed to automatically sort the references
\usepackage{tabu}                      % only used for the table example
\usepackage{booktabs}                  % only used for the table example
\usepackage[table]{xcolor}
\usepackage{graphics}
\definecolor{c1}{rgb}{0,0.3,1}
\definecolor{c2}{rgb}{1,0,0.0}
\definecolor{c3}{rgb}{0.16, 0.5, 0.0}
\definecolor{c4}{rgb}{0.5, 0.5, 1}

\newcolumntype{?}{!{\vrule width 1pt}}

\onlineid{0}

\vgtccategory{Research}
\vgtcpapertype{application/design study}

\title{Multiscale Snapshots: Visual Analysis of Temporal Summaries in Dynamic Graphs}

\author{Eren Cakmak, Udo Schlegel, Dominik J\"ackle, Daniel Keim, and Tobias Schreck
}
\authorfooter{
\item Eren Cakmak, Udo Schlegel, and Daniel Keim are with the University of Konstanz, Germany. E-mail: firstname.lastname@uni-konstanz.de 
\item Dominik J\"ackle is an Independent Researcher, Munich, Germany. \\ E-Mail: dominikjaeckle@gmail.com
\item Tobias Schreck is with the TU Graz, Austria. \\ E-Mail: tobias.schreck@cgv.tugraz.at
}

\shortauthortitle{Cakmak \MakeLowercase{\textit{et al.}}: Multiscale Snapshots: Visual Analysis of Temporal Summaries in Dynamic Graphs}
\abstract{The overview-driven visual analysis of large-scale dynamic graphs poses a major challenge. 
We propose Multiscale Snapshots, a visual analytics approach to analyze temporal summaries of dynamic graphs at multiple temporal scales. 
First, we recursively generate temporal summaries to abstract overlapping sequences of graphs into compact snapshots. 
Second, we apply graph embeddings to the snapshots to learn low-dimensional representations of each sequence of graphs to speed up specific analytical tasks (e.g., similarity search). 
Third, we visualize the evolving data from a coarse to fine-granular snapshots to semi-automatically analyze temporal states, trends, and outliers. 
The approach enables us to discover similar temporal summaries (e.g., reoccurring states), reduces the temporal data to speed up automatic analysis, and to explore both structural and temporal properties of a dynamic graph. 
We demonstrate the usefulness of our approach by a quantitative evaluation and the application to a real-world dataset. 
} % end of abstract


\teaser{
  \centering
  \includegraphics[width=\linewidth]{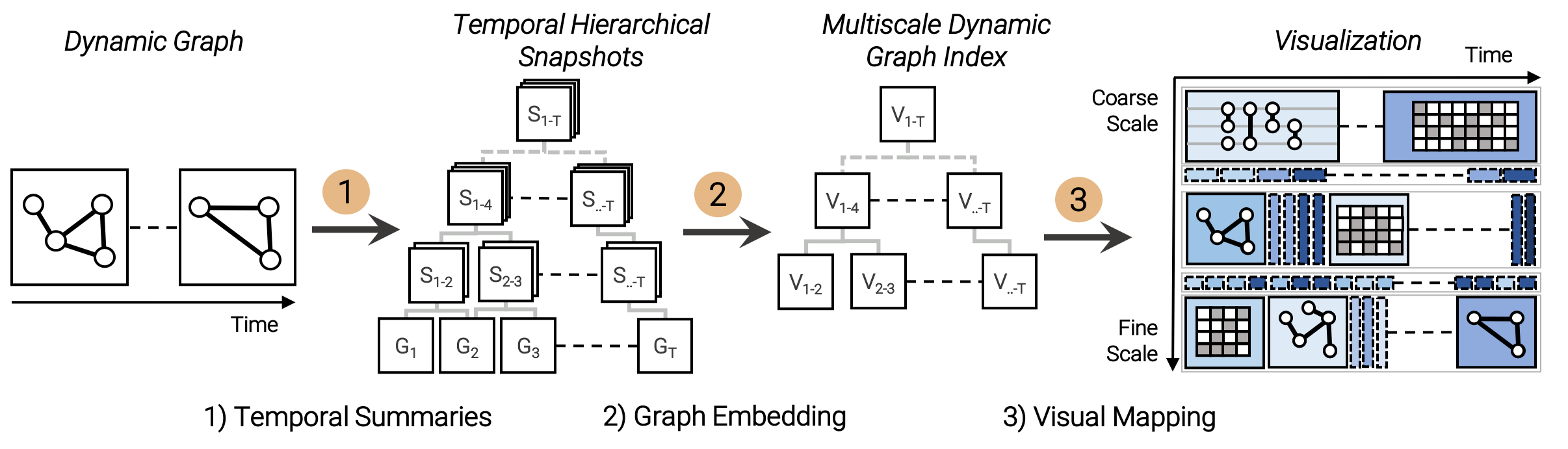}
  \vspace{-0.6cm}
  \caption{
    Overview of our Multiscale Snapshots approach: We (1) recursively create temporal summaries (snapshots) of graphs at different temporal scales (time granularities); 
    (2) We then apply an unsupervised graph learning method (graph embedding) to learn low-dimensional representations of snapshots;
    Then, we (3) then give an exploratory visualization that organizes the snapshots of different temporal granularities in a hierarchy, to provide an overview of the evolving structural properties, which utilizes the graph embeddings for analytical tasks (e.g., similarity search). 
    The approach allows rich interaction with the temporal and structural aggregates to understand large time-dependent graphs, for example, to visually analyze reoccurring temporal states. 
  }
  \label{fig:teaser}
}

\vgtcinsertpkg

\begin{document}

%% The ``\maketitle'' command must be the first command after the
%% ``\begin{document}'' command. It prepares and prints the title block.

%% the only exception to this rule is the \firstsection command
\firstsection{Introduction}

\maketitle
%% the only exception to this rule is the \firstsection command
%%%%%%%%%%%%%%%%%%%%%%%%%%%%%%%%%%%%%%%%%%%%%%%%%%%%%%%%%%%%%%%%%%%%%%%%%%%
% INTRODUCTION
%%%%%%%%%%%%%%%%%%%%%%%%%%%%%%%%%%%%%%%%%%%%%%%%%%%%%%%%%%%%%%%%%%%%%%%%%%%
% \section{Introduction} \label{sec:intro}
% FIRST PARAGRAPH
% - what is the main characteristic of the data with examples
% - what is the KEY TASK!
% - what is the main problem/challenge regarding the task
A dynamic graph models changing relationships between entities over time. 
Many real-world data analysis problems rely on dynamic graphs,  including, among others, social, computer, and communication networks, and in practice, contain large amounts of dynamic data, hence presenting challenges for effective exploration.
% - Main Task
An important task in such dynamic graphs is to obtain an overview of the evolving topology by identifying meaningful temporal intervals and their underlying changing structural properties~\cite{burch2017dynamic}.
% Examples 
For instance, analysts are often interested in the identification of stable, reoccurring, transition, and outlier states~\cite{van2016reducing}.
% - Main Challenge(s)
However, as dynamic graphs are often large-scale and evolve over long periods, it is a major challenge to identify suitable analysis methods and present the data in a readable, scalable, and expressive manner~\cite{beck2017taxonomy}. 
% Temporal abstraction
Previous approaches for visual analysis of dynamic graph data, therefore, often incorporate temporal abstraction methods (e.g., temporal aggregation and dimensionality reduction) to provide an overview of higher-level structures over time~\cite{van2016reducing}. 
In real-world applications, the usefulness of such temporal abstraction methods depends on many factors, including the selection of an appropriate temporal scale, the user task at hand, graph size, and frequency of topological changes.
% Solution 
Currently, visual analytics systems for dynamic graphs lack methods for the visual analysis of dynamic processes at different temporal abstraction scales (multiscale analysis), often leaving the analyst with the challenging task of distinguishing overlapping temporal changes manually.

%%
% THIRD PARAGRAPH
% - what do we propose - high level
% - what is the research question
% - how does our approach work, what do we apply
% - what are the advantages of our approach
% - application examples and use cases we show to demonstrate the usefulness
% What
We propose \emph{Multiscale Snapshots}, a novel visual analytics approach to semi-automatically provide a multiscale overview of structural and temporal changes in dynamic graphs. 
% How 
We combine temporal hierarchical abstractions with unsupervised graph learning methods to enable the identification of similar evolving graphs. 
% The temporal hierarchy - Why
First, the temporal hierarchical snapshots summarize the dynamic graph recursively at multiple temporal scales to reduce the size of the large-scale data.
% Graph embedding - why
Second, we apply unsupervised graph learning (e.g., graph2vec~\cite{narayanan2017graph2vec}) to the snapshots of the hierarchy to learn low-dimensional representations of graph sequences, which enables users to use the embeddings for analytical tasks (e.g., similarity search) and later on to adapt the temporal scale semi-automatically. 
% The visualization - why
Third, the visualization of the hierarchy of snapshots provides an overview of trends, allows users to compare periods, and to explore structural as well as temporal properties of dynamic graphs.
% Why useful 
The approach enables exploring various abstraction methods at multiple temporal scales to provide an overview of large dynamic graphs temporal and structural properties.

% Overall results
With Multiscale Snapshots, we can retrace how dynamic patterns and changing graph properties affect the overall evolving data, and compare temporal structures at different levels of temporal resolution. 
The contributions of this work are the following: 
(1) The Multiscale Snapshots approach to visually analyze temporal and structural similarities at multiple temporal scales; 
(2) A temporal hierarchical abstraction using unsupervised graph learning methods to reduce the size of dynamic graphs and speed up analytical tasks (e.g., similarity search).

The remainder of this paper is structured as follows. 
Section 2 discusses the related work. 
Section 3 describes the addressed problems, the research gap, and the design goals.
In Section 4, we describe the Multiscale Snapshots approach and our implementation. 
In Section 5, we quantitatively evaluate and apply the approach to a real-world, large-scale dynamic graph. 
A discussion is given in Section 6, before Section 7 concludes. 

%%%%%%%%%%%%%%%%%%%%%%%%%%%%%%%%%%%%%%%%%%%%%%%%%%%%%%%%%%%%%%%%%%%%%%%%%%%
% RELATED WORK
%%%%%%%%%%%%%%%%%%%%%%%%%%%%%%%%%%%%%%%%%%%%%%%%%%%%%%%%%%%%%%%%%%%%%%%%%%%
\section{Related Work} \label{sec:related_work}
Multiscale Snapshots combines temporal summaries with graph embeddings to present an overview of the underlying dynamic phenomena.
% What comes next
In the following, we discuss related work from automated analysis, visualization, visual analytics, and multiscale visualization approaches for dynamic graphs.

%%% --------------------------------------
%%% -------- New subsection --------------
\subsection{Dynamic Graph Analysis and Visualization}
% - Before jumping into a category, first sentence should be a Topic Sentence which defines the obvious gaps
% \looseness=-1
The visual analysis of long graph sequences has lately gained research attention~\cite{beck2017taxonomy}.
% -- Automated analysis methods  
The automatic analysis, such as the temporal analysis of static as well as dynamic graph metrics (e.g., centrality, diameter~\cite{brandes2004analysis}, or change centrality~\cite{federico2012visual}), enables us to examine structural properties of the data.
% Graph embeddings 
Furthermore, recent approaches in unsupervised learning focus on embedding graph structures into low-dimensional space~\cite{zhang2018network}.
% What is the problem here 
However, only analyzing such automatically extracted structural properties might hide specific local dynamic changes (e.g., changes of density) and may fail to capture the overall dynamic phenomena~\cite{zhang2018network}. 
% Solution - interactive vis - investigate network metrics in structural evolving 
Interactive visualizations can overcome these challenges by allowing analysts to explore the dynamic relationships in their evolving structural context,
% Visualization methods 
and several visualization techniques for dynamic graphs have been proposed.
Popular approaches display dynamic graphs as animations~\cite{diehl2002graphs, purchase2006important, archambault2011animation}, timeline~\cite{hadlak2013supporting, cui2014let, bach2015time, dal2017wavelet, xu2018exploring} and hybrid visualizations~\cite{rufiange2013diffani, bach2015small, burch2015flip}. 
For further reading, we refer to the surveys of Kerracher et al.~\cite{kerracher2015task}, Beck et al.~\cite{beck2017taxonomy}, and Nobre et al.\cite{nobre2019state}. 

% Limitations 
Many of the proposed visualization techniques, however, do not scale to a large number of nodes, edges, and time steps at the same time~\cite{hadlak2011situ}. 
% Bridge to VA approaches 
Consequently, to adapt existing techniques to large-scale dynamic graphs, visual analytics approaches were proposed that combine automatic analysis methods with interactive visualizations to reduce the presented data and highlight structural changes.

%%% --------------------------------------
%%% -------- New subsection --------------
\subsection{Visual Analytics of Dynamic Graphs}
% - Before jumping into a category, first sentence should be a Topic Sentence which defines the obvious gaps
The visual analytics of dynamic graphs aims to seamlessly integrate graph analysis methods~\cite{newman2018networks} with visualization techniques~\cite{beck2017taxonomy} to interactively analyze the evolving structural properties.
Such visual analytics approaches facilitate abstraction methods for large-scale dynamic graphs to reduce the amount of data and provide an overview of high-level changes.
% What is the problem here 
The usefulness of such abstraction methods, however, depends on the applied method (e.g., temporal clustering) and input parameters (e.g., number of clusters)~\cite{hadlak2013supporting}.
% Solution 
Therefore, according to Aigner et al.~\cite{aigner2008visual}, it is essential to interactively adapt abstraction methods and tune their underlying parameters to identify changes that otherwise would remain hidden.

% Data abstraction methods for dynamic networks in summary 
In general, there are two categories of abstraction methods; \textit{data space abstraction} (e.g., sampling, clustering) and \textit{visual space abstraction} (e.g., zooming, focus-and-context)~\cite{cui2006measuring}.
% - General topic of data abstraction methods  
The \textit{data space abstraction} in dynamic graphs reduces the number of graph elements or time steps~\cite{schulz2013grooming}. 
Often, data space abstraction methods lower the resolution of the data (e.g., temporal aggregation~\cite{moody2004dynamic}).
% VA approach of segmenting and aggregating  
For instance, Van den Elzen et al.~\cite{van2016reducing} proposed a visual analytics approach that segments and aggregates sequences of graphs to a vector and applies dimensionality reduction to obtain an overview of the states in dynamic graphs.
However, the resulting overview depends strongly on the selected segmentation scale and the abstraction method (extracted features) into vectors.
Further, the dimensionality reduction technique is hard to interpret as the projection does not visualize the evolving graph structure.
% Further reading 
For an overview of \textit{data space abstraction} methods, we recommend the recent survey of Liu et al.~\cite{liu2018graph}.
% Visual abstraction methods for dynamic networks 
% - General topic of visual abstraction methods 
The \textit{visual space abstraction} methods in dynamic graphs reduce the amount of presented data (e.g., by applying zooming~\cite{bederson1994pad++}). 
Many of the \textit{visual space abstraction} methods allow the user to interactively change the depicted level of detail~\cite{abello2004matrix,henry2006matrixexplorer,archambault2008grouseflocks,elmqvist2008zame,wong2009multi,bezerianos2010graphdice,zinsmaier2012interactive}.
% Temporal navigation methods 
For example, temporal navigation methods help to interactively adapt the horizontal (e.g., TempoVis~\cite{ahn2011temporal}) and vertical (e.g., Bender-deMoll and McFarland~\cite{bender2006art}) time dimension. 
Multiple visual analytics approaches, including \textit{visual abstraction methods}, were recently proposed.
For instance, Small MultiPiles~\cite{bach2015small} enables users to interactively stack and present a sequence of graphs as piles of adjacency matrices to reduces the number of displayed views.
% Cubix 
Furthermore, Cubix~\cite{bach2014visualizing} allows users to visually explore adjacency matrices of dynamic graphs in a cube metaphor.
However, in both approaches, the identification of temporal patterns in long sequences of adjacency matrix visualizations remains challenging due to limited display space and overlapping issues in 3D visualizations.

% Bridge to segmentation and 
Many of the previously proposed approaches focus mainly on aggregation and display the abstracted temporal or structural dimension at one scale, which makes it challenging to investigate the influences of abstraction methods on the resulting visualization, as patterns may be found at different scales and intervals. 
Multiscale visualizations aim to overcome these challenges by simultaneously displaying different levels of abstraction, hence providing an encompassing overview of possible structural and temporal aggregation levels.

%%% --------------------------------------
%%% -------- New subsection --------------
\subsection{Multiscale Dynamic Graph Visualizations}
% - Before jumping into a category, first sentence should be a Topic Sentence which defines the obvious gaps
Multiscale (multiresolution) visualizations present the data at multiple user-defined levels of abstraction and are useful to set detailed abstraction levels into the overall temporal context~\cite{elmqvist2010hierarchical}. 
For example, Javed and Elmqvist~\cite{javed2012stack} stack different levels of zoomed time-series data in a tree structure to serve as a graphical history and preserve the context during zooming.
% Not many proposed methods 
Nearly all of the previously mentioned approaches visualize the dynamic graphs on a single time granularity (scale) using mostly one adjustable abstraction method.
% Only a few of the already proposed approaches present a dynamic graph simultaneously at multiple scales. 
One notable exception is the recent work of Burch and Reinhardt~\cite{burch2017dynamic} that proposed a timeline visualization technique that allows exploring dynamic graphs at different temporal granularities.
However, the approach focuses on bipartite graphs, and due to the overplotting produced by the interleaving method, the identification of temporal patterns remains challenging.

% \subsection{Delineation to our Work}
% Summary 
Most of the listed approaches for dynamic graphs focus on the analysis of dynamic graphs at a particular temporal scale and require the manual definition of parameters. 
For instance, the work of Van den Elzen et al.~\cite{van2016reducing} requires users to define a discretization scale,  feature selection, and the choice of a suitable dimensionality reduction technique. 
% The problems of all of the previous approaches
In contrast to these approaches, we propose using temporal hierarchical abstractions with unsupervised learning methods to explore input parameters (e.g., discretization scale) and simultaneously visualize graph sequences at different levels of temporal abstraction.

%%%%%%%%%%%%%%%%%%%%%%%%%%%%%%%%%%%%%%%%%%%%%%%%%%%%%%%%%%%%%%%%%%%%%%%%%%%
% Basic Idea
%%%%%%%%%%%%%%%%%%%%%%%%%%%%%%%%%%%%%%%%%%%%%%%%%%%%%%%%%%%%%%%%%%%%%%%%%%%
\section{Application Background} \label{sec:problem}
% Why 
The analytical goal of our approach is to provide an overview of evolving graph properties at multiple abstraction scales.
% How 
In the following, we describe the addressed problems, the research gap we close, and our derived design goals.

\begin{table}[tb]
  % \caption{VIS/VisWeek accepted/presented papers: 1990--2016.}
  \scriptsize%
	\centering%
	\resizebox{\columnwidth}{!}{%
  \begin{tabular}{|r?c|c|c?c|c|c?c|c|c|c|c|c|}
  \hline
    \textbf{Publication, Year} & \multicolumn{3}{|c|}{\textbf{Visualization}}  & \multicolumn{3}{|c|}{\textbf{Scalability}} & \multicolumn{6}{|c|}{\textbf{Temporal Explorability}}\\ \specialrule{.1em}{.05em}{.05em} 
    
    & \rotatebox{90}{Animation \vspace{0.1cm}} 
    & \rotatebox{90}{Timeline} 
    & \rotatebox{90}{Hybrid} 
    & \rotatebox{90}{Small }%($<$ 1000)} 
    & \rotatebox{90}{Large }%($>$ 1000)} 
    & \rotatebox{90}{Multiscale} 
    & \rotatebox{90}{Nodes} 
    & \rotatebox{90}{Edges}  
    & \rotatebox{90}{Neighbors}   
    & \rotatebox{90}{Paths} 
    & \rotatebox{90}{Clusters}
    & \rotatebox{90}{Subgraphs}\\ \specialrule{.1em}{.05em}{.05em} 

    \rowcolor{gray!50} \textbf{Multiscale Snapshots} & & & $\bullet$ &  $\bullet$ & $\bullet$ & $\bullet$ & $\bullet$ & $\bullet$ &  $\bullet$ & & $\bullet$ & $\bullet$ \\  \specialrule{.1em}{.05em}{.05em} 
    %    \multicolumn{13}{|c|}{} \\ \hline
    Rufiange \& McGuffin~\cite{rufiange2013diffani}, 2013 & & & $\bullet$ & $\bullet$ & & & $\bullet$ & $\bullet$ & $\bullet$ & $\bullet$ & &\\ \hline
    Hadlak et al.~\cite{hadlak2013supporting}, 2013 & & $\bullet$ & & $\bullet$ & &  $\bullet$ & & & & & $\bullet$ & $\bullet$ \\ \hline
    Cui et al.~\cite{cui2014let}, 2014 & & $\bullet$ & & $\bullet$ & $\bullet$ & & & & $\bullet$ & & $\bullet$ &\\ \hline
    Bach et al.~\cite{bach2015small}, 2015 & & & $\bullet$ & $\bullet$ & & & & & & & $\bullet$ &  $\bullet$\\ \hline
    Burch \& Weiskopf~\cite{burch2015flip}, 2015 & & & $\bullet$ & $\bullet$ & & & $\bullet$ & $\bullet$ & $\bullet$ & & $\bullet$ &\\ \hline
    Bach et al.~\cite{bach2015time}, 2015 & & $\bullet$ & & $\bullet$ & $\bullet$ & $\bullet$ & & & & & &\\ \hline
    Van den Elzen et al.~\cite{van2016reducing}, 2016 & & $\bullet$ & & $\bullet$ & $\bullet$ & & $\bullet$ & $\bullet$ & $\bullet$ & & &\\ \hline
    Burch et al.~\cite{burch2017dynamic}, 2017 & & $\bullet$ & & $\bullet$ & & $\bullet$ & & $\bullet$ & & & $\bullet$ &\\ \hline
    Dal et al.~\cite{dal2017wavelet}, 2017 & & $\bullet$ & & $\bullet$ & & & $\bullet$ & & $\bullet$ & & &\\ \hline
    Xu et al.~\cite{xu2018exploring}, 2018 & & $\bullet$ & & $\bullet$ & & & $\bullet$ & $\bullet$ & & & $\bullet$ &  $\bullet$\\ \hline
    Wang et al.~\cite{wang2019nonuniform}, 2019 & & $\bullet$ & & $\bullet$ & & & $\bullet$ & $\bullet$ &  $\bullet$  &  $\bullet$  &  &  \\ \hline
  \end{tabular}%
}
 \vspace{0.1cm}
 \caption{
    The comparison highlights the essential temporal properties of related visualization techniques, ordered by publication date, and assessed by us to the best of our knowledge from analysis of the works. 
    The \textbf{visualization} category classifies the techniques using the taxonomy of Beck et al.~\cite{beck2017taxonomy}. 
    The \textbf{scalability} category elaborates on multiscale temporal approaches and the temporal scalability with large scalability meaning a dynamic graph with more than 1000 graphs. 
    The \textbf{temporal explorability} is adapted from the work of Nobre et al.~\cite{nobre2019state} and illustrates whether the graph structures (e.g., neighbors, clusters) are explorable and comparable within the temporal dimension.}
    \vspace{-0.7cm}
     \label{tab:comparison}

\end{table}
%% ---
\subsection{Problem Description}
% Overview visualization
\looseness=-1
The starting point of data analysis is often an overview visualization to examine the overall data structure, and to identify useful analytical and visualization techniques~\cite{shneiderman2003eyes}.
% Dynamic graphs 
However, providing an overview of large-scale dynamic graphs can be challenging for multiple reasons~\cite{beck2017taxonomy}.
% First 
First, the complexity and size of data pose various challenges as many dynamic graph visualization techniques do not scale~\cite{beck2009towards}.
% Second 
Second, it is challenging to visualize dynamic graphs as there is a trade-off between displaying the detailed graph structure for each time step, and also presenting the evolving properties over time.
For instance, animations support the exploration of each static graph over time. 
However, animations are considered unsuited to provide an overview of long periods due to the problems caused by cognitive effort~\cite{tversky2002animation,healey2012attention,archambault2011animation} and difficulties maintaining a mental map in dynamic graphs~\cite{purchase2006important}. 
% Third 
Third, creating compact temporal abstractions (summaries) of dynamic graphs is user task-dependent and relies upon the application domain as well as data properties.
% Example scale 
For example, a fine-grained temporal aggregation in large-scale dynamic graphs results in various intervals with little information unable to provide an overview~\cite{bender2006art}. 
In contrast, coarse-scale aggregation produces only a few intervals, which may contain a high variance, where meaningful intervals could go unnoticed.
Finding appropriate levels of abstraction is a non-trivial task~\cite{kerren2007human}.
% Main challenge 

\subsection{Gaps in Related Approaches}
\looseness=-1
We compare a selection (see Table~\ref{tab:comparison}) of recent work on dynamic graph visualization to point out the gap we intend to close.
% How selected 
The selected publications are based on the recursive search of references from the recent surveys of Beck et al.~\cite{beck2017taxonomy} and Nobre et al.~\cite{nobre2019state}.
The categories of our comparison comprise visualization techniques from the dynamic graph taxonomy~\cite{beck2017taxonomy}, the temporal scalability, including multiscale approaches, and the temporal explorability of different graph structures~\cite{nobre2019state}.

The comparison reveals several insights. 
% Multiscale vis 
First, the number of temporal multiscale approaches for dynamic graphs is limited.
% Time series 
Multiscale approaches present either time series in a multiresolution design (e.g., graph metrics~\cite{hadlak2013supporting}) or include visualizations having multiscale characteristics (e.g., time curves~\cite{bach2015time}).
% Timeline vis  
Second, timeline visualizations (time-to-space mappings) reduce the size of the data and are suitable to provide an overview of a sequence of graphs. 
However, such approaches abstract and discretize structural information at one temporal scale (uniform time slicing), often requiring users to manually identify overlapping temporal patterns~\cite{wang2019nonuniform}.
% Example 
For example, timeline-based visualizations often require to define input parameters, e.g., discretization parameters and derived features~\cite{van2016reducing}.
% Complex vis 
Third, current work either introduce a single complex visualization, or the combination of various simple views, which poses the question of how to arrange the different visualizations in the limited screen space and link them effectively.

% Summary 
In summary, a significant number of visualization techniques view a dynamic graph as a series of static graphs which neglects to simultaneously capture the evolving structural properties of dynamic graphs. 
% \us{why?}
% What we do 
In contrast to previous approaches, we interactively apply an unsupervised graph learning method (graph embeddings) on a multiscale temporal hierarchy to directly learn structural properties.
We furthermore use graph embeddings as a familiar representation for an analytical user task (e.g., similarity search, comparison) and utilize the visual exploration of different visual metaphors. 
% \ts{appropriate (?) may sound better than simple} 

\subsection{Design Goals}
% Design goals
We derived three design goals for our visual design from the previously described problem description, the research gap, and also research challenges outlined in related work~\cite{beck2009towards,archambault2014temporal,beck2017taxonomy}.

\textbf{G1: Time-oriented visual analysis} 
% Why 
The visual analysis of dynamic graphs lacks new paradigms to examine structural (static) and temporal properties simultaneously. 
% How 
First, the identification of structural properties, e.g., clusters in a static graph, allows searching for similar structural properties over time. 
% Questions 
Such an exploration enables users to identify temporal states and continue to search for similar trends, reoccurring structures, and outlier structures. 

\textbf{G2: Temporal multiscale overview} 
% Why 
Our core idea is to provide an overview of multiple levels of temporal granularity, which facilitates users to relate higher-level overviews with low-level details.
% How 
Such a multiscale overview allows detecting useful temporal analysis scales and gives additional context while navigating the temporal dimension (e.g., temporal filtering). 
For instance, a multiscale overview allows comparing states and transitions across multiple temporal granularities.

\textbf{G3: Multiple visual metaphors} 
The combination of different visual metaphors in a consistent interface allows adjusting the visualization to the data characteristics of particular intervals. It also increases the task coverage by enabling an analyst to adapt the visual representations to the task at hand.
% Example 
For example, matrix-based visualization techniques are better suited for dense subsequences of dynamic graphs.

%%%%%%%%%%%%%%%%%%%%%%%%%%%%%%%%%%%%%%%%%%%%%%%%%%%%%%%%%%%%%%%%%%%%%%%%%%%
% Multiscale Hybrid Visualization
%%%%%%%%%%%%%%%%%%%%%%%%%%%%%%%%%%%%%%%%%%%%%%%%%%%%%%%%%%%%%%%%%%%%%%%%%%%
\section{Multiscale Snapshots} \label{sec:multiscale_visualization}
% Multiscale Snapshots 
Multiscale Snapshots provides an overview of higher-level and fine-grained temporal intervals of large-scale dynamic graphs.
% Basic idea 
The approach reduces the complexity of the data by integrating temporal summarization and graph embeddings in an interactive multiscale visualization.

% VA approach 
Our proposed visual analytics approach consists of three adjustable steps (see Fig.~\ref{fig:teaser}) to promote the exploration and summarization tasks for dynamic graphs~\cite{brehmer2013multi}.
% Temporal summaries
The first step transforms the temporal dimension into a hierarchy of snapshots summarizing subsets of the dynamic graph into overlapping multiscale intervals.
% Vector generation
The second step reduces the complexity of the snapshots by embedding the summaries of the evolving topology into vector representations (e.g., using graph2vec~\cite{narayanan2017graph2vec}).
The mapping of intervals into vector representation allows us to automatically adjust the visualization to highlight temporal states, trends, and outliers.
% Visual mapping 
The third step transforms the abstracted temporal data into a flexible and interactive hierarchical visualization and supports essential interaction as well as navigation methods to analyze the evolving graph structure visually.
% The hierarchy 
Furthermore, the visualization intends to increase the task coverage by combining different visualizations of dynamic graphs in a consistent interface.
The following subsections describe each transformation step in more detail. 

\subsection{Temporal Hierarchical Snapshots}
\begin{figure*}[tb]
 \centering 
 \includegraphics[width=\linewidth]{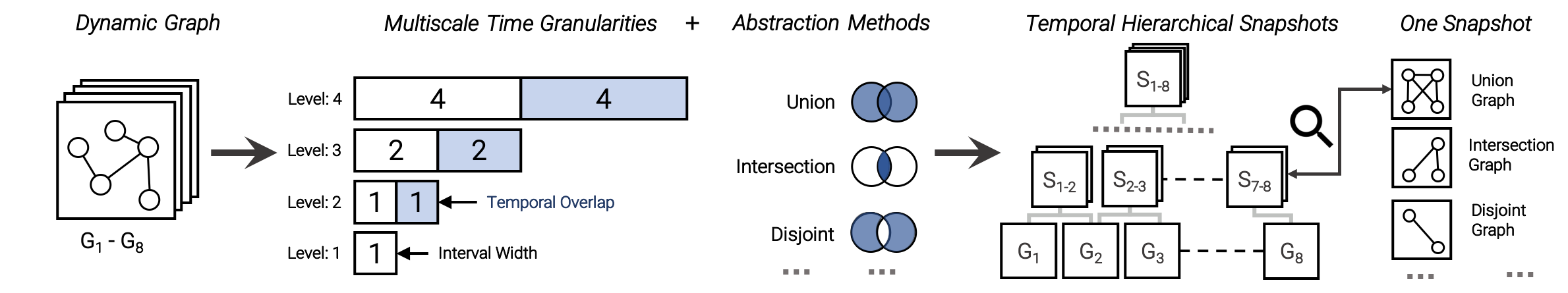}
 \vspace{-0.4cm}
 \caption{
 The Figure displays the generation of temporal hierarchical snapshots for a dynamic graph with eight timesteps. 
 %\dominik{add g1 to g8 to the first dynamic graph to make it very obvious that you are talking about the eight time steps and thus, eight graphs. Of is the first image already the aggregated snapshot?} 
 First, the dynamic graph is partitioned into overlapping intervals at four levels of temporal granularity. 
 The fourth level contains all data, and the first level consists of intervals of size one (static graphs). 
 Second, abstraction methods are applied to the intervals to generate different compact summaries of the subsets of the dynamic graph.
 The result is a hierarchy of temporal snapshots that contains multiple summary graphs (e.g., union, intersection, and disjoint graph. }
 \vspace{-0.4cm}
 \label{fig:temporal-granularity}
\end{figure*}
% Temporal data
Dynamic graphs model relationships over time (e.g., social networks) and can be described as a number of $T$ static graphs $DG = (G_1, G_2, ... ,G_T)$.
% Graph 
% A graph for the timestep $i$ defined as $G_i = (V_i, E_i)$ consisting of a set of nodes (vertices) $V_i$ and a set of edges with $E_i \subseteq V_i \times V_i$. 
% Temporal abstraction
The temporal abstraction of dynamic graphs (e.g., aggregation, filtering) helps to reduce the data size, speed up temporal queries, support interactive analysis, and eliminates noise~\cite{liu2018graph}.
% Challenges
However, the temporal abstraction of sequence graphs into summaries remains challenging due to the selection of \textit{time granularity}, which depends on many factors (e.g., application, data size), as well as the choice of \textit{abstraction method} (e.g., summarization).

% Time slicing 
In various dynamic graph visualizations, a simple selection of one \textit{time granularity} (uniform time-slicing) is used due to the simplicity of the approach~\cite{wang2019nonuniform}. 
% Time granularity 
In contrast, we propose a recursive temporal abstraction into a hierarchy with temporal overlaps to model multiple \textit{time granularities} (see Fig.~\ref{fig:temporal-granularity}).
% How 
We generate and stack multiple partitionings using uniform intervals (time slices) of different temporal granularities.
% Hierarchy 
We organize the stacked partitionings in a hierarchy that orders the different levels of abstraction (discretizations) from coarse to fine-grained temporal representations.
% Goal 
Our bottom-up approach groups per default the temporal dimension into intervals of length $2^l$ with the level $l \in 1,..., \lceil log(T) \rceil$.
% Example 
Fig.~\ref{fig:temporal-granularity} displays an example partitioning for a dynamic graph with eight time steps.
% Level 0 
The level one of the hierarchy consists of intervals of length one, containing only one graph of the evolving data. 
The intervals are generated using a rolling window method, which facilitates time discretization without hard boundaries. 
The rolling window approach for level $l$ is computed by shifting the interval of width $2^{l}$ by the temporal overlap of width $2^{l-1}$.
This results in each level having $\lceil T / 2^{l-1} \rceil$ intervals and the whole hierarchy having $(3 \cdot T) - 1$ intervals.
% Summary 
Essentially, as seen in Fig~\ref{fig:temporal-granularity}, each generated interval overlaps partly (e.g., per default half) with the next interval except for the level one (single graph) and the root node (all graphs).
% Height 
The default recursive partitioning into multiscale intervals results in the height of $\lceil log(T) \rceil$.
% Alternativen
In practice, for most datasets, the height of the hierarchy is below 20 ($<$ 1 million graphs).
% Alternative parameters 
The width of time slicing can be modified to the application domain, for example, intervals with a width of a day, week, month, year.

% Other suggestion  
The uniform time-slicing produces intervals of the same width for each level. 
The generation of non-uniform intervals for each level can be computed by applying temporal clustering techniques with varying input parameters.
% \dominik{This is not only an \textbf{also} - rather it is a solution to also having non-uniform intervals, because reviewers will on it for sure.}
For example, the temporal clustering approach of Hadlack et al.~\cite{hadlak2013supporting} can be used to identify similar substructures based on graph properties to provide an overview of temporal trends.
% Unsupervised methods 
A hierarchy of temporal intervals can also be automatically generated by facilitating unsupervised learning with boundary detectors to obtain hierarchical temporal dependencies at different time scales~\cite{chung2019hierarchical}.
The generation of such hierarchical temporal dependencies only works on time-series data of a dynamic graph, for example, on evolving graph metrics such as average clustering coefficient or density.
Therefore, applying such methods remains challenging as there is no single graph metric that can capture all of the evolving graph structures. 

% Why graph abstraction 
The \textit{temporal abstraction} aims to summarize and capture the evolving structural properties of sequences of graphs. 
% Representative 
We suggest utilizing multiple abstraction methods to generate diverse representations of the generated multiscale intervals as there is not a single abstraction method able to encode all evolving properties of a dynamic graph.
%\dominik{why multiple? I suppose the reason is that a single one does not meet the requirements?}
% How 
We transform the intervals into graph summaries per default using set operations (union, intersection, disjoint graph). 
% Example 
For example, the union operation abstracts the interval into a supergraph, which helps to provide an overview of all nodes and edges~\cite{hadlak2011situ}.
% Three summaries 
The three default graph summarization techniques (see Fig.~\ref{fig:temporal-granularity}) are the \textit{union graph} that consists of a union of the set of all node and edge sets. 
The \textit{intersection graph} which consists of all nodes appearing more than $i$-times in the interval.
The \textit{disjoint graph} consists of all nodes appearing less than $i$-times in the interval.
%\ts{Can you argue why this setting? Based on tests, or a  conceptual reason?} 
We set the default value for the parameter $i$ to the interval overlap of an interval.
If the values of $i$ are below the interval overlap, this will most probably result in successive intervals with a similar intersection and disjoint graphs.

% Snapshots 
We call the three computed graph summaries of an interval a snapshot $S_{l,k}$ (see Fig.~\ref{fig:temporal-granularity}).
A snapshot aims to capture the structural and temporal properties of a sequence of graphs on level $l$ and the $k$ generated interval.
The resulting intervals of the snapshots can be indexed in an interval tree to support the efficient support window queries, for example, identifying the best fitting interval to a user-defined period.
% Suggestion 
We suggest, furthermore, utilizing more graph summarization methods based on the analytical task, data characteristics, and application domain.
% Example 
For example, we implemented a community detection algorithm (e.g., the Clauset-Newman-Moore algorithm~\cite{clauset2004finding}) to reduce the overall number of nodes in each static graph and to extract higher-level properties (e.g., meta-nodes and edges).
% More info 
For more graph summarization methods that can be added to our approach, we refer to the survey by Liu et al.~\cite{liu2018graph}.

% Summary
Overall, the first step results in a hierarchy of abstracted snapshots at different temporal granularities (see Fig.~\ref{fig:temporal-granularity}).
Every interval in the hierarchy contains multiple graph summaries, which can be used for different types of queries later on. 
For example, we can search for similar changes between intervals by using the disjoint graphs to identify reoccurring changes in the dynamic graph.
The resulting temporal hierarchy of the dynamic graphs is used in the next step of our Multiscale Snapshots approach by mapping each summary to its vector representation.

\subsection{Multiscale Dynamic Graph Index}
% Why 
As for the next step, the resulting hierarchical snapshots are learned and embedded into low-dimensional space to reduce the complexity of the graphs and speed up analytical tasks (e.g., similarity search). 
% Goal 
The main goal is to use unsupervised learning methods to model the similarities between the different multiscale temporal summaries and reduce the complex data characteristics to low-dimensional vectors preserving information.
% How
We apply a graph embedding (e.g., graph2vec~\cite{narayanan2017graph2vec}) to map all snapshot graphs (e.g., union and disjoint graphs) to vector representations.
% Advantage 
In contrast to earlier approaches (e.g., Van den Elzen~\cite{van2016reducing}), unsupervised graph learning methods learn the topological structures of graphs and do not require any hand-engineered features.
% Index 
The embeddings can be precomputed and are also typically small enough to fit into main memory.
% First VA approach 
To the best of our knowledge, Multiscale Snapshots is the first visual analytics approach to propose using unsupervised graph learning methods with different temporal granularities to visually analyze intervals sharing similar properties over time.

% Graph embeddings 
Recently, new unsupervised graph learning methods have been proposed to learn node and graph embeddings~\cite{zhang2018network}. 
% Static 
However, many of these methods mainly focus on learning static graph embeddings and cannot model the evolving properties of dynamic graphs~\cite{zhang2018network}. 
% The suggestion
In contrast to earlier approaches, we propose to model dynamic graphs by embedding summaries of subsets of the evolving data to capture the temporal dependencies between graphs.
% How 
The analyst can apply graph embeddings such as graph2vec~\cite{narayanan2017graph2vec}, GL2Vec~\cite{chen2019gl2vec}, and FGSD~\cite{verma2017hunt} to the snapshots.
% Suggestion 
The approach embeds all snapshots of the temporal hierarchy except for the level one (single graphs), which results in the embedding of $2T-1$ snapshots.
% Dynamic graph index 
The resulting $2T-1$ embeddings are also indexed to support efficient K-nearest neighbor search queries. 
% What exactly
We employ the following two index structures: an interval tree to support efficient temporal queries for the intervals, and an individual index structure for each level. 
We utilize for the indexing of the graph embeddings the proposed method of Malkov et al.~\cite{malkov2018efficient} to perform a fast K-nearest neighbor search in each level.

% Evaluation 
In our evaluation (see Sec.~\ref{sec:evaluation}), we compare different unsupervised graph embeddings, discuss the scalability of the approaches, and show that the embeddings of the snapshots are able to capture structural as well as temporal changes. % \us{sounds weak but needs to be adapted later}

%%-------
\subsection{ Multiscale Snapshots Visualization}
\begin{figure}[tb]
 \centering 
 \includegraphics[width=\linewidth]{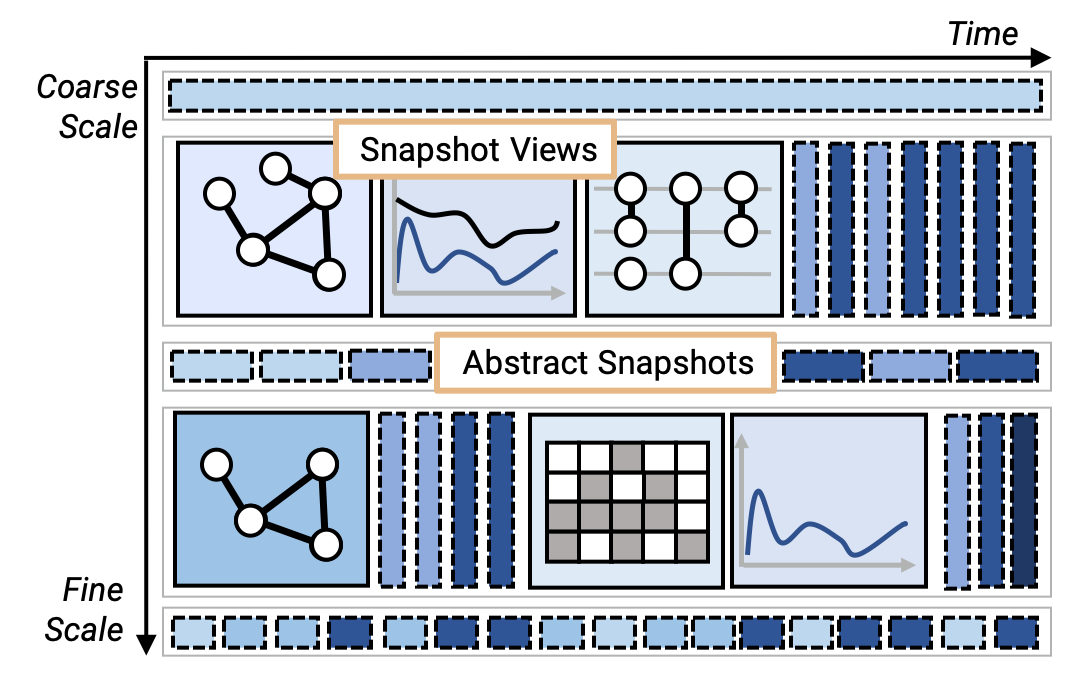}
 \vspace{-0.4cm}
 \caption{The hierarchy organizes and displays the summaries from the snapshots from coarse to fine-grained representations.
  The visual metaphors in each snapshot view can be manually or semi-automatically adapted. 
  The snapshot views can be abstracted to reduce the number of displayed views and duplicate information.
  The background color of each snapshot is mapped to graph metrics (e.g., number of edges).
  }
  \vspace{-0.4cm}
 \label{fig:snapshot-view}
\end{figure}

The final step applies a visual mapping to organize the temporal snapshots in a multiscale visualization to enable the visual analysis of the generated snapshots and uses the graph embeddings for analytical tasks.
% Design rationales
In the following, we describe the components of our visual and interaction design (see Fig.~\ref{fig:snapshot-view}). 

The visualization presents the hierarchy of snapshots and orders them from coarse to granular scale (top-down) and facilitates the horizontal (time) as well as vertical (time granularity) temporal navigation to search for similar properties over time (\textbf{G1}).
% Why a hierarchy 
The visualization stacks and displays the multiscale temporal abstractions (\textbf{G2}), allowing to analyze and compare the abstracted data at different temporal granularities.
Presenting multiple abstraction levels enables us to gain more knowledge about the underlying abstracted dynamic graph (e.g., data distribution)~\cite{elmqvist2010hierarchical}. 
% How
The highest level (root) displays an aggregated version of the whole dynamic graph (e.g., union graph), and the bottom level enables us to depict a limited number of each time step.
The levels in-between allow visualizing a subset of the generated snapshots in snapshots views (juxtaposed small multiples).

% Why 
A snapshot view combines different visual metaphors in a consistent interface to increase the task coverage (\textbf{G3}) and displays one of the summary graphs (e.g., union graph).
Every view enables users to depict the data using four kinds of visual metaphors (node-link, adjacency matrix, animation, and time series of graph metrics).
% Why these four visualizations 
We use these visual metaphors since the individual benefits, and drawbacks of the representations are well studied (e.g., graph layout and matrix reordering)~\cite{beck2017taxonomy}.
% Why multiple visual metaphors
We utilize multiple visual metaphors for certain intervals as the usefulness of dynamic graph visualization depends on the underlying changing data (e.g., sparse versus dense graphs)~\cite{burch2017dynamic}.
% What  
We consider our snapshot views as hybrid visualizations, as the view combines different visual metaphors in small multiple representations.
% Graph size 
Furthermore, the Clauset-Newman-Moore community detection algorithm~\cite{clauset2004finding} is applied to minimize visual clutter and to reduce the number of nodes in a snapshot view, if the size of the displayed summary graph exceeds a specific threshold (more than 100 nodes). 
This threshold is based on the size classification of Nobre et al.~\cite{nobre2019state}.
% Resulting in meta nodes 
The resulting communities are then shown as meta-nodes and allow to filter the respective nodes and edges of the community for the entire Multiscale Snapshot visualization. 
For instance, the filtering of a structural cluster allows us to explore the evolving properties of the cluster in the displayed snapshot views. 
% Background color
The snapshot view also visualizes derived structural properties using the background color of each snapshot view to highlight differences between adjacent visual metaphors.
% Graph metrics 
The derived properties (graph metrics) of the summary graph are used to identify and emphasize temporal or structural graph properties.
% Example 
For instance, we compute graph metrics such as the sum of the number of edges in a snapshot, which indicates the density of the underlying graph sequence.
A linear color scale from light blue (low values) to darker blue (high values) is used to highlight changes of the derived structural properties~\cite{harrower2003colorbrewer}.

% Drawbacks 
The use of multiple levels of juxtaposed small multiples remains challenging due to limited display space and the preservation of the viewer's mental map.
% Drawbacks 
The simultaneous presentation of multiple levels and their snapshot views does not visually scale due to the restricted display space with an increasing number of snapshot views, as the readability of each view decreases.
% Solution 
We, therefore, incorporate visual space abstraction methods to limit the number of displayed levels and snapshots views. 
% For example,
The number of displayed levels is limited (default four), and during the vertical navigation, the respective lowest or highest level of temporal granularity is removed.
% Abstractions 
Furthermore, we abstract snapshot views to reduce the number of shown visualizations and on particular snapshots while keeping the context of the abstracted views (focus-and-context principle).
An abstracted snapshot is displayed as a compact colored rectangle without any visual representation.
% Which methods 
The background color can be mapped to extracted graph metrics of the selected summary graph, for example, the number of nodes as well as edges, average clustering coefficient, density, and transitivity.
The coloring of such abstracted snapshot views enables the identification of intervals with specific properties, such as subsequences of dense graphs.
% Why 
In general, the usage of such color indicating graph properties allows users to identify and compare temporal intervals~\cite{tominski2008task}. 
The abstraction can be done manually by reducing individual snapshot views or whole levels of the hierarchy, using a user-driven threshold, and an automated abstraction algorithm. 

% Automated algorithm
The automated algorithm limits the number of intervals by traversing the hierarchy and abstracting redundant information. 
The algorithm abstracts snapshots if the number of views exceeded a specific threshold, or if the algorithm detects duplicate displayed periods.  
% How 
The algorithm traverses each level of the hierarchy (top-down) and compares the displayed snapshots at each level against each other. 
If coarse snapshots (high level) are displayed in the fine-grained snapshots (low levels), they are abstracted.
% How 
The automatic abstraction is done based on overlapping windows in the interval tree, which means that the snapshot view with the highest overlap with low-level snapshots is abstracted.
% % How 
The algorithm compares, for example, the time interval of the root view against all other not abstracted snapshots, and if the periods of these more granular levels display the majority of temporal information of the root view, then the root snapshot view is abstracted.
% Threshold 
The thresholds for the automatic abstraction algorithm, such as the overall number of levels and snapshot views, are adjustable by the analyst.

Furthermore, we aim to preserve the viewer's mental map, which increases the readability and interpretability of the evolving data~\cite{purchase2006important}.
% Mental map problem and layout  
To maintain the viewer's mental map, we fix and use one global layout for each visual metaphor.
% Example
For instance, we compute one layout for the overall supergraph of the dynamic graph.
% Still challenging 
Furthermore, the usage of linking and brushing aims to preserve the mental map between adjacent snapshots that are using different visual metaphors and the different levels of abstraction in the hierarchy. 

% - Solution - VA
Multiscale Snapshots utilizes the graph embeddings for automated analysis to identify trends, reoccurring, and outlier states. 
% Example 
For example, an analyst can select a snapshot view and can apply a k-nearest neighbor search query to detect similar summary graphs (see the query interface Fig.~\ref{fig:use-case-full}). 
% Further analysis 
The detected k-nearest neighbor snapshots can also be disaggregated to more granular views using the interval tree (drill-down).
% Selection 
The similarity search can also be applied to a particular type of summary graph, for instance, search for similar intersection graphs.
% Advantage 
Such similarity queries also enable to semi-automatically abstract and adapt the displayed snapshot views. 
% How 
The k-nearest neighbor queries can also be applied to particular intervals (subqueries) and to specific levels, which allows examining the summaries of the dynamic graphs in a top-down manner. 
% Furthermore
The embeddings can also be used to cluster levels of the hierarchy and to identify outlier states by applying outlier detection algorithms~\cite{aggarwal2015outlier}. 

% Summary 
% Why 
In summary, the visual design provides an overview of snapshots of a dynamic graph by combining automatic analysis methods with visual space abstraction methods (focus-and-context).

%%%%%%%%%%%%%%%%%%%%%%%%%%%%%%%%%%%%%%%%%%%%%%%%%%%%%%%%%%%%%%%%%%%%%%%%%%%
% Visual Analysis of Dynamic Networks based on Multiscale Snapshots
%%%%%%%%%%%%%%%%%%%%%%%%%%%%%%%%%%%%%%%%%%%%%%%%%%%%%%%%%%%%%%%%%%%%%%%%%%%
\subsection{Multiscale Snapshots Prototype}
\label{sec:prototype}
% Goal 
We show the usefulness of our approach by applying it to real-world data using our prototype~\footnote{\href{https://github.com/eren-ck/MultiscaleSnapshots}{https://github.com/eren-ck/MultiscaleSnapshots}}. 
% How 
The prototype consists of two components (see Fig.~\ref{fig:use-case-full} A-B): the Multiscale Snapshots visualization and the query interface.
Both components allow users to explore and semi-automatically search for similar temporal states in the dynamic graph.

% Visualization 
The \textbf{Multiscale Snapshots visualization} consists of a toolbar, the stacked snapshot views, and two context bars.
The toolbar facilitates the application of automated analysis methods (e.g., open the query interface) and visualizes the summary graphs of the snapshots (e.g., display union or intersection graph).
Furthermore, the toolbar enables changing the data space abstraction methods (e.g., filter and cluster nodes) and adapting visual transformations (e.g., reordering algorithms for matrix visualization).
The prototype displays by default the root of the hierarchy as a supergraph (union graph) using a node-link diagram visualization.
% Layout 
The layout of the node-link diagram is computed once for the root supergraph using per default the Fruchterman-Reingold~\cite{fruchterman1991graph} layout algorithm and later used for all snapshot views.
% Start 
The hierarchy enables an analyst to navigate horizontally (time) or vertically (overview to detail) on the temporal dimension.
% Why context bars 
The two context bars display additional information during the horizontal and vertical navigation of the temporal dimension. 
% What 
The time context bar on the top shows the visualized intervals, and the level context bar on the right allows to add and remove levels.
% Snapshot views
Each snapshot view can be visually analyzed via zooming, panning, brushing, and changing the layout in all views (e.g., reordering of the matrix) to make visual patterns more apparent~\cite{behrisch2016matrix}.
% Visual metaphor 
The visual transformations for individual or all snapshot views can be adjusted by the analyst to enable the adaption of visual metaphors to the underlying sequence of graphs, such as switching for periods of dense sequences of graphs to matrix visualization.
The prototype also enables filtering by specific graph properties (e.g., node degree) and clustering~\cite{clauset2004finding} to reduce overall displayed elements to extract higher-level features (e.g., meta-nodes and edges).
The background color of each snapshot view can be mapped to extracted graph metrics and computed node characteristics (e.g., clustering coefficient) to node size.
% How 
To apply a k-nearest neighbor query, an analyst has to select a specific summary graph in a snapshot view. 

% Why query interface 
The \textbf{query interface} allows applying specific k-nearest-neighbor queries to search for similar summary graphs on all or particular levels of temporal granularity.
% Why 
The query interface displays each time dimension of a level and encodes the currently visualized and already investigated snapshots via color. 
This additional information helps to keep an overview of the already explored snapshots of all levels.
% Sorted 
The timelines can be ordered by different features, such as by the percentage of explored snapshots. 
% Filter and selection 
An analyst can select the levels, time interval, and the summary graph type (e.g., only union graphs) to apply the k-nearest neighbor search.
% K-nearest neighbors 
The number of k-nearest neighbors is also configurable. 
% Query 
The query results are displayed as dots on the timeline, and the euclidean distance between the underlying graph embeddings is mapped to the opacity of the dot.
The analyst has to select a subset of the nearest-neighbors, which are then displayed in the Multiscale Snapshots visualization.
% Result 
The selected results are shown as snapshot views and allow users to analyze and compare similar temporal states in lower or higher temporal granularities against each other.

%%%%%%%%%%%%%%%%%%%%%%%%%%%%%%%%%%%%%%%%%%%%%%%%%%%%%%%%%%%%%%%%%%%%%%%%%%%
% Use Case
%%%%%%%%%%%%%%%%%%%%%%%%%%%%%%%%%%%%%%%%%%%%%%%%%%%%%%%%%%%%%%%%%%%%%%%%%%%
\section{Evaluation}\label{sec:evaluation}
% Why 
In the following section, we evaluate the two main components of the Multiscale Snapshots approach.
% Visualization 
We provide a usage scenario to demonstrate how the visual analytics approach can be utilized to gain an overview of temporal summaries in a dynamic graph.
% How 
We furthermore quantitatively evaluate the similarity ($k$-nearest neighbors) search of the graph embeddings with synthetic and real-world datasets.

%% -----
% Usage Scenario 
\subsection{Usage Scenario}
\label{sec:use-case}
\begin{figure*}[tb]
 \centering 
 \includegraphics[width=\linewidth]{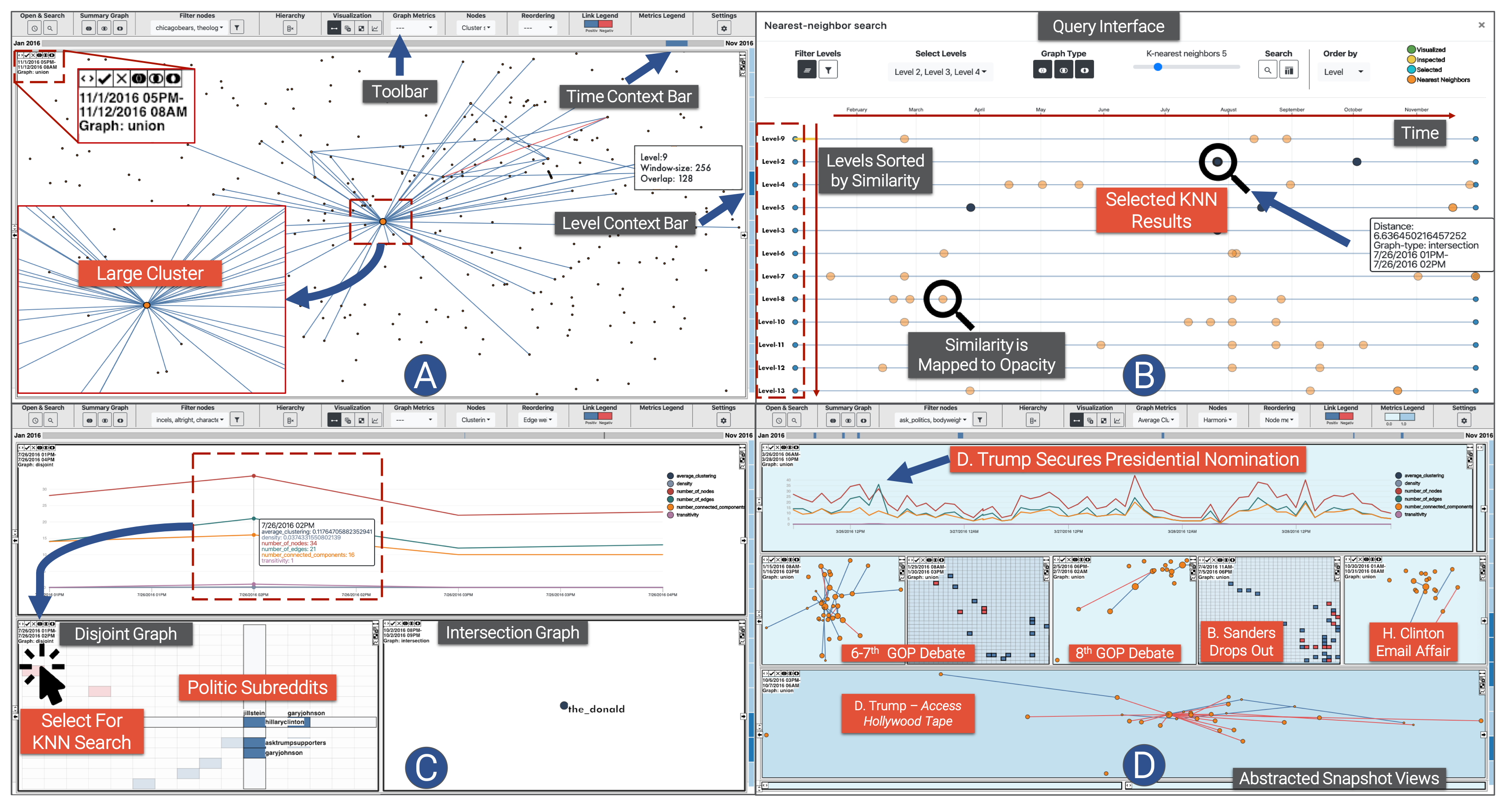}
 \vspace{-0.6cm}
 \caption{
 The prototype implementation consists of two primary components the Multiscale Snapshots visualization (A) and the query interface (B). 
 The figures present the visual analysis of the Reddit hyperlink dataset (see Sec.~\ref{sec:use-case}).
 % The described visual analysis of the Reddit hyperlink dataset (see Sec.~\ref{sec:use-case}).
 The displayed nodes are subreddits, and the edges are timestamped hyperlinks between subreddits with either positive (blue) or negative (red) sentiment.
 The displayed nodes are subreddits, and the edges are timestamped hyperlinks between subreddits with either positive (blue) or negative (red) sentiment.
 The example illustrates by the case of the 2016 US election how the approach allows searching for similar temporal states in the dynamic graph.
 The intermediate steps of the visual analysis and the resulting interfaces are presented in the sub-figures C-D.
 In D, the results of the visual analysis by similarity search are displayed, which are significant events in the timeline of the presidential election. 
 }
 \vspace{-0.4cm}
 \label{fig:use-case-full}
\end{figure*}
% Why
We demonstrate the applicability of our approach using a real-world, large-scale dynamic graph of the website Reddit~\cite{kumar2018community}.
% How 
Reddit is a social news aggregation website with 440 million active users who can publish and upvote posts of interest (e.g., link to news sites) in particular communities (subreddits).
% What 
% In each subreddit, users can post content (e.g., links to news sites), and the community then decides via a voting based system which posts are interesting.
% Dataset 
The analyzed dataset is a dynamic hyperlink graph in which nodes are subreddits, and edges are hyperlinks posted between subreddits. 

% Analysis task 
\textbf{Tasks}
In the following, we outline the actions that a fictitious analyst takes to discover structural and temporal changes during the 2016 US presidential elections (see Fig.~\ref{fig:use-case-full}).
% Tasks
A task in the visual analysis of such hyperlinks is to gain an overview of temporal events (e.g., political scandals), identify reoccurring links between communities, and examine structural changes within groups of subreddits.
% Why it is not possible with state of the art systems 
The visual analysis of such data with state-of-the-art visual analytics approaches remains challenging due to the varying duration of such events. 
For example, the length of political scandals varies significantly due to media exposure and their temporal context (e.g., during elections).
% Multiscale analysis 
In contrast to previous approaches, Multiscale Snapshots enables us to detect events/states of different temporal lengths due to the temporal multiscale modeling.

% Sentiment 
\textbf{Dynamic Graph}
% Dataset 
The Reddit dataset~\cite{kumar2018community} comprises hyperlinks between subreddits from the 1st January 2016 to 30th November 2016.
% Statistics 
The resulting dynamic graph contains 7974 graphs (grouped by hours), 18546 subreddits (nodes), and 88328 hyperlinks between subreddits (edges).
% Edges 
The timestamped hyperlink posts have a sentiment label indicating if the post is positive or negative towards the other subreddit.
% The embeddings 
The dynamic graph index was computed using the Graph2Vec~\cite{narayanan2017graph2vec} embedding approach for 80 epochs, and three summary graphs for each snapshot were generated (union, intersection, and disjoint graphs).
% Validation 
The validation of the detected findings is done by comparison to the ground truth of real historical news coverage.

\looseness=-1
\textbf{Initial Setup} 
%Per default, our prototype displays the entire dynamic graph as an aggregated node-link diagram (supergraph) and computes a global layout using the Fruchterman-Reingold algorithm~\cite{fruchterman1991graph} once for all snapshot views which aims to preserve the mental map during the visual analysis.
Per default, our prototype displays the entire graph as an aggregated node-link diagram (supergraph). 
Then, based on the Kamada-Kawai algorithm~\cite{kamada1989algorithm}, a global layout is computed for all snapshot views once. 
This way, the mental map is preserved during the visual analysis.
% Clustering 
Furthermore, snapshot views that display more than 100 nodes are automatically clustered using the greedy Clauset-Newman-Moore community detection algorithm~\cite{clauset2004finding} to reduce the number of nodes and to extract higher-level properties (e.g., meta-nodes and meta-edges).
% Example 
The clustering of the approximately 8000 nodes of the analyzed data reveals several clusters of subreddits (e.g., computer games subreddits).

% Election week 
\textbf{Starting Point: Election Week} 
First, the analyst wants to analyze the election week of the 2016 US presidential race (8th November 2016) to identify important groups of political subreddits.
% View
The analyst enters the dates of the election week, and the prototype automatically searches for the best fitting snapshot period using the interval tree.
% Union graph 
The prototype depicts a union graph of election week, and the analyst maps the size of the cluster to the node size to discover large groups of subreddits (see Fig.~\ref{fig:use-case-full}-A). 
% Display the large cluster 
He selects the largest visible meta-node and all underlying political subreddits of the cluster.
% Select and filter 
He filters these political subreddits as he assumes that the political subreddits of the election week have also been active in the political discourse of the whole election.

% Search for similar events 
\textbf{Similarity Search} 
To identify political events similar to the election week in the dynamic graph, the analyst searches for similar embeddings using the election weeks supergraph. 
% Nearest neighbor query 
Using the query interface (see Fig.~\ref{fig:use-case-full}-B), he queries the five nearest neighbors for each level and sorts the levels by the similarity of the embeddings.
% How 
The executed nearest neighbor query is calculated on the unfiltered summary graphs for each snapshot, which means that the similarity search results will include false-positives that do not necessarily include any political subreddits.
% Result 
The analyst discovers that the results of the query are similar embeddings on the second (2-hour periods) and third level (4-hour periods), which means that these rather short sequences of graphs consist of a subset of hyperlinks similar to the ones during the election week.
% interpretation 
The analyst selects the three closest neighbors for both levels and therefore navigates from a high temporal aggregation (a week) to a lower granularity (2-4 hours).
% Similarity search 
Three queried snapshots are empty, meaning the views do not contain any of the previously filtered political subreddits.
% Empty results 
The empty views are presumably false positives that capture other graph sub-structures of the election week.
% Removing empty snapshots 
The analyst removes the three empty snapshots and examines the remaining three snapshots by changing the visual mapping from a node-link diagram to the time series of graph metrics. 

% Three snapshots 
\textbf{Fine-Grained Temporal Analysis} 
The three remaining snapshots contain a different amount of nodes. 
The intersection graph on the second level contains only one subreddit (\textit{the$\_$donald}), which means the subreddit was referenced in both graphs of the snapshot (2 hours).
% High level - low level includes 
The analyst discovers that a high-level summary graph (disjoint) includes a displayed snapshot of the second level.
The unexpected overlap steers the analysts towards the low-level disjoint graph, which seems to also be the peak in the time series of graph metrics.
The time series presents the number of nodes, edges, as well as connected graph components, the graph density, the average clustering coefficient, and the transitivity over time.
It seems that this second level snapshot is essential for the results of the search as the snapshot has similar graph structures compared to the supergraph of the election week.
% Peak 
The peak is presented as a matrix visualization (see Fig.~\ref{fig:use-case-full}-C) and can be attributed to the events of the national democratic convention where H. Clinton was nominated for the presidential election.
% The matrix 
The disjoint graph represented as a matrix visualization (see Fig.~\ref{fig:use-case-full}-C) can be associated to the political event of the democratic nomination H. Clinton which resulted in a cluster of hyperlinks between political subreddits (e.g., \textit{hillaryclinton}, \textit{asktrumpsupporters}, and \textit{garyjohnson}) and other hyperlinks between political subreddits (e.g., \textit{communism101}, \textit{altright}, and \textit{crazyideas})
% Knn search
The analyst uses the snapshot (disjoint graph) for another similarity search. 
He expects the similarity search to return more political events because the low-level graph embedding of the two-hour snapshot contains mainly linked political subreddits.

% The result 
\textbf{Searching for Political Events} 
The similarity search finds many similar snapshots at different temporal granularities, which indicates that these political events also seem to be discussed for different periods.
% What 
The query returns several similar snapshots of the sixth level with an interval length of 32 hours, which can refer to potential political events and their daily news coverage scheme (see Fig.~\ref{fig:use-case-full}-D).
% The analyst 
The analyst investigates the different snapshot views, mostly union, and disjoint graphs, and abstracts all snapshot views with only a few subreddits.
% The remaining snapshot views 
The remaining presented snapshots are on levels 5-7 and contain intervals of 16, 32, and 64 hours.
% The analyst 
The analyst maps the average clustering coefficient to the background color of each snapshot view to identify periods with structural clusters.
% The dense periods 
He changes the visual metaphors of the dense snapshots to matrix visualizations and the higher-level periods to the time-series metaphor.
% What
The different visual metaphors allow the analyst to put the events on lower levels into the overall temporal context, for example, the analyst can relate how the linkage behavior between subreddits declines after political scandals.

% Results 
\textbf{Political Events and Scandals} 
The analyst then visually analyzes the periods and sees that during the selected periods political subreddits link each other, mainly in a positive (blue edge) or negative (red edge) way.
% Example 
Various subreddits such as \textit{the$\_$donald} and \textit{asktrumpsupporters} usually have positive hyperlinks between each other.
% The timeline 
He examines external resources of the timeline of major events for the 2016 US elections and can refer the presented snapshot views to events in the presidential race.
% Example 
The sixth level of the hierarchy displays several GOP (republican party) political debates, B. Sanders dropping out of the primary election, and the H. Clinton Email affair.
% B. Sanders 
The analyst is also able to identify structural changes between the views, for instance, after B.\ Sanders drops the linking activity of some subreddits (e.g., \textit{SandersForPresident} or \textit{Democratic Socialism}) declines. 
% Billy bush 
The snapshot view of 6-7th October on the fifth level stands out as it mostly contains negative links between the subreddits. 
The analyst can relate the period to the leaked tapes of the 2005 Access Hollywood show in which D.\ Trump brags about sexual exploits and also on the same day WikiLeaks published the email of H.\ Clintons campaign manager revealing her paid Wall Street speeches.
% Animation 
The analyst wants to analyze this snapshot further and displays the supergraph as an animated node-link diagram to examine the spread of news between the subreddits on an hourly basis.
% More stuff 
During the further analysis of snapshot views, the analyst can also detect other events, for instance, the final nomination of D.\ Trump by the GOP, which results in visible changes in the time series plot of graph metrics.
% But 
He also detects some events, which he cannot relate to actual major political events directly.
% Interpretation 
Those events are probably general political discussions initiated by Reddit users or targeted distribution of news from public-relations groups or political bots.
% How to check this 
To further investigate such events, the analyst can select these non-assignable events and search for similar periods, for example, to identify the reoccurring post of political bots.

%% -----
% Evaluation 
\subsection{Experimental Evaluation}
\begin{table*}[tb]
  \scriptsize%
	\centering%
	\resizebox{\linewidth}{!}{
 \begin{tabu}{|r|[1.25pt]c|c|c|c|c|[1.25pt]c|c|c|c|c|[1.25pt]c|c|c|c|c|[1.25pt]}
\hline
 
 & \multicolumn5{|c|}{Synthetic Data} 
 & \multicolumn5{|c|}{Reddit Data~\cite{kumar2018community}} 
 & \multicolumn5{|c|}{Wikipedia Data~\cite{paranjape2017motifs}} 
 % & \multicolumn4{|c|}{Dataset 1} &  \multicolumn4{|c|}{Dataset 2} &  
 \\ \hline  \hline
  \textbf{Interval length}  
    & 1 & 2 & 3 & 4 & 8 
    & 1 & 2 & 3 & 4 & 8 
    & 1 & 2 & 3 & 4 & 8  \\ \hline  \hline
 
    Graph2Vec 
    & 0.096 &	0.146&	0.096&	0.15&   0.096  % Synthetic data 
    & 0.332 &	0.066& 	0& 	0& 	0.264  % Reddit 
    & 0 & 0 & 0 & 0 & 0.066 % Wiki
    \\ \hline
    
    GL2Vec 
    & 0.096&	0.122&	0.024&	0.024&	0 % Synthetic data 
    & 0.198&  	0.132&  	\textbf{0.266} & 	0.066&  	0.066  % Reddit 
    & 0.066 & 	0 & 	0.066 & 	0 & 	0.066  % Wiki
    \\ \hline
    
    FGSD 
    & 0.146 &	\textbf{0.224}&	\textbf{0.198}&	\textbf{0.198}&	0.048 % Synthetic data 
    & \textbf{0.464} & 	\textbf{0.332} & 	0& 	\textbf{0.132}& 	0.066 % Reddit 
    & \textbf{0.866} &	\textbf{0.132} &	0.066&	0.132&	0.134 % Wiki
    \\ \hline
    
    Multiscale Graph2Vec 
    & \textbf{0.148}&	0.072&	0.096&	0.072&	\textbf{0.122} % Synthetic data 
    & 0.266& 	0.264& 	0& 	0.066& 	0.264 % Reddit
    & 0.234&	0.066&	0.066&	0.066&	0.198 % Wiki
    \\ \hline
    
    Multiscale GL2Vec 
    & \textbf{0.148}&	0.072&	0.096&	0.072&	\textbf{0.122} % Synthetic data 
    & 0.198& 	0.132& 	0.066& 	\textbf{0.132} &	\textbf{0.4} % Reddit 
    & 0.234 & 	0.066& 	0.198& 	0.132& 	0.198 % Wiki
    \\ \hline
    
    Multiscale FGSD 
    & 0.146&	0.096&	0.096&	0.072&	\textbf{0.122} % Synthetic data
    & 0.264& 	0.264& 	0& 	0& 	0.332 % Reddit 
    & 0.466&	0.066&	\textbf{0.2}&	\textbf{0.198}&	\textbf{0.266} % Wiki
    \\ \hline
\end{tabu}%
}
\vspace{0.1cm}
 \caption{
     % Why 
    The Table presents the quantitative evaluation results of the k-nearest neighbor search with and without the multiscale graph index for different graph embedding methods. 
    % Values 
    The average accuracy values for window queries of different lengths (1-8) are depicted for each dataset. 
    % What 
    The experiment was repeated five times on synthetically generated data and with randomly selected subsets of real-world data.
    % Results 
    The results of the evaluation indicate an improved accuracy on window queries on the listed real-world dataset.
 % \us{Quantitative evaluation of our multiscale index on k-nearest neighbor interval search on synthetic and real-world datasets. We compare the results of three embedding techniques with and without using multiscale index.}
 }
  \label{tab:results}

    \vspace{-0.7cm}
\end{table*}

The generated graph embeddings for the multiscale snapshots are independent of any analytical task and can be used for clustering, graph prediction, and outlier detection.
% Graph embeddings 
In the following, we show that the multiscale graph embeddings allow us to search for similar sequences of graphs.
% What 
Across all experiments, we use the same parameter settings for the generation of the multiscale index. 
% Consequence

\textbf{Problem Background} 
A similarity search for a set of graphs can be interpreted as a query to return $k$-nearest neighbors to a specific graph.
% Brute force solution 
An exhaustive simple brute-force algorithm would compute the distance between all graphs, for example, the graph editing distance (GED)~\cite{bunke1983distance} and return the list of $k$ nearest graphs.
% Extensive search 
However, the extensive brute-force approach does not scale as the GED computation is not feasible for graphs with more than 16 nodes~\cite{blumenthal2018exact}.
% Faster solutions 
Therefore, heuristics are usually applied to decrease the computation effort of $k$-nearest neighbor queries, which in turn frequently reduces the accuracy of the results.
% Interval query 
In the following, we apply window queries for sequences of graphs to show that summarization methods (e.g., union graph) can capture some temporal characteristics.

\textbf{Datasets} 
% Why 
We evaluate the performance of similarity search on synthetic and real-world data.
% How 
We generated five synthetic dynamic graphs using the dynamic stochastic block model with diminishing communities~\cite{goyal2018dynamicgem}.
% What exactly 
The synthetic datasets consist of 150 nodes with three communities and 100-time steps, containing varying amounts of diminishing communities (up to 20-time steps) in which two nodes are exchanged for each time step.
% Real-world datasets 
We evaluate the approach with real-world datasets. 
% Reddit data 
The Reddit data~\cite{kumar2018community} is a dynamic hyperlink graph with subreddits (nodes) and hyperlinks or crossposts between subreddits (edges).
% Wiki data
The Wikipedia dataset~\cite{paranjape2017motifs} consists of a dynamic graph that captures the editing behavior (edge) between Wikipedia Talk pages (nodes).
% Group by hour 
For each real-world dataset, we preprocess the data by computing a supergraph for each hour, which generates descriptive dynamic graphs with more than two nodes per time step.
% Picking a subset 
We evaluated our approach on randomly picked subsets (100 graphs) of the real-world data. 
We select a subset of the data as the computation of the following ground truth is quite expensive.

\textbf{Ground Truth}
% How 
We calculate a ground-truth similarity score for the $k$-nearest neighbor search by computing the distance between the input and all other graphs.
% How 
We employ the following similarity measure between two graphs.
Our similarity measurement first models the graphs as two adjacency matrices $A$ and $B$ and then compute for each matrix the singular values via the singular value decomposition.
Afterward, we calculate the $fnorm$ using 
\[ {fnorm} = \sqrt{\sum_{i=0}^S \sigma_i^2} \]
We define the distance between two graphs as
\[ {madist}(A, B) = |{fnorm}(A) - {fnorm}(B)| \]
Using the given similarity measurement, we compute the distances between all graphs to obtain a ground-truth of $k$-nearest neighbors. 

\textbf{Baseline Methods} 
% Why 
We used three unsupervised graph learning methods on the described datasets. 
The graph embeddings are applied once with and once without the multiscale temporal modeling. 
We used the following graph embedding methods~\cite{karateclub} with the described input parameters:
\begin{itemize}
    \setlength\itemsep{0.1cm}
    \item \textbf{graph2vec}~\cite{narayanan2017graph2vec}: 250 epochs, 0.025 learning rate, 2 Weisfeiler-Lehman iterations, and 128 dimensions.
    \item \textbf{GL2Vec}~\cite{chen2019gl2vec}: 250 epochs, 0.025 learning rate, and 128 dimensions.
    \item \textbf{FGSD}~\cite{verma2017hunt}: 200 number of histogram bins with a the histogram range of 20.
\end{itemize}
% Multiscale snapshot 
For window queries for the single graph embeddings without any summarization methods, we utilize the median value of the embeddings as the representative value of the interval. 
% Why Median 
We use the median as the average of the embeddings as these can lead to potential distortions in the embedded space.
% Multiscale Snapshots 
For the multiscale temporal embedding, we applied only one temporal summarization method to generate a union graph for each snapshot.
% How 
The searched intervals for the $k$ nearest neighbor search are extracted before training of embedding techniques.
% How this is done 
We randomly extracted five intervals with different lengths ($1-8$) from the dynamic graph and randomly removed one node from each graph.

\textbf{Evaluation Metrics}
The following metrics are used to evaluate the approach.
% MSE 
We compute the accuracy of the $5$-nearest neighbor queries based on the ground-truth. 
For the accuracy computation, we do not incorporate the ordering of the nearest neighbors and expect only the presence in the result set.

\textbf{Experimental Setup}
All experiments were computed on a computer with two CPU cores (Intel i7-6567U 3.30GHz) and 16 GB RAM.
The experiment was repeated five times, and the average accuracy was computed for each randomly picked interval with different lengths.

\textbf{Results}
The results are described in Table~\ref{tab:results}. 
% FGSD 
The results indicate that FGSD~\cite{verma2017hunt} works best to identify nearest neighbors on an embedding basis using the median.
The results show that the single graph embeddings have a higher accuracy on the synthetic data.
In contrast, the real-world datasets indicate different results by demonstrating equal or improved results by using the multiscale index for longer intervals ($<4$).
% Explanation
An explanation for this can be the fact that there is a drastic difference between the topology of the synthetic and real-world datasets.
% Real world data
For example, in the real-world data nodes and edges are added and removed more frequently between time steps.
The synthetically generated dataset has a quite high density, while in contrast to this, the real-world datasets are much more sparse.
% Synthetic data 
For example, in the synthetic data, the nodes are just moved between the clusters, so only edges change over time.
These synthetic properties prevent supergraphs from encoding the topological changes over time.
% Solution 
Therefore, the multiscale graph index requires a different temporal summarization method to capture the changes of the synthetic dataset (e.g., disjoint graph). 

%%%%%%%%%%%%%%%%%%%%%%%%%%%%%%%%%%%%%%%%%%%%%%%%%%%%%%%%%%%%%%%%%%%%%%%%%%%
% Discussion
%%%%%%%%%%%%%%%%%%%%%%%%%%%%%%%%%%%%%%%%%%%%%%%%%%%%%%%%%%%%%%%%%%%%%%%%%%%
\section{Discussion} \label{sec:discussion}
% Multiscale Snapshots - why and how
The Multiscale Snapshots approach consists of three steps: (1) applying temporal summarization methods, (2) utilizing graph embedding methods to reduce the size of the generated graph summaries, and (3) the visual analysis of the generated snapshots.
% - Interpretation of Evaluation/User Study
% Why 
Our quantitative evaluation indicates the usefulness of the multiscale graph embeddings, and the usage scenario shows the application of the approach to real-world data.
% Discussion - which methods should apply? 
Overall the utility of the approach yet depends on multiple aspects (e.g., summarization method and graph embedding), the data characteristics (e.g., data distribution), and the task at hand (e.g., outlier analysis).

Steps (1-2) involve multiple methods with parameters. For instance, the graph embeddings methods require a definition of the number of layers and epochs.
% Input parameters
For an analyst, such parameter choices pose a challenge as he has to determine suitable methods and their input parameters to generate useful embeddings. 
% Positive 
We consider the flexibility of using different temporal summarization methods and graph embeddings as an advantage of our approach and a possibility for future work. 
% Example 

% Computational scalability 
Another challenge for steps (1-2) is the computational scalability for the precomputation of the embeddings.
% Large scale graphs
For example, the computation of a dynamic graph of length $T$ with $|V|$ nodes and $|E|$ edges require for only union graphs $O( log(T) \cdot (|V|+|E|) )$ memory and time complexity.
% Computational scalability 
We speed up the computation of temporal summaries by parallelizing each levels snapshot generation and the usage of an interval tree.
% Graph embeddings 
Goyal and Ferrara~\cite{goyal2018graph} surveyed the time complexity of graph embeddings, and scalable embeddings run in the time complexity of $O(|E|)$.
% Server
Due to the time and memory complexities, we suggest computing the graph embeddings for large scale dynamic graphs on a server.
% Example 

% -  Limitations and Opportunities (related to design choices)
% - Limitations that do not hurt too much and could be opportunities
% Why 
Step (3) aims to display the temporal dimension at multiple scales, which poses new user-related aesthetic challenges~\cite{beck2009towards}.
% - Mental map 
% Why 
To preserve the mental map, we compute and use only one layout for each applied visual metaphor (e.g., global node-link diagram layout).
% How 
An analyst can change the global layout for all snapshot views, for instance, by reordering the cells and rows of the adjacency matrix visualization.
% Another challenge 
The snapshot views can also result in adjacent snapshots that display different dynamic graph visualization (e.g., node-link and matrix visualization).
% Why 
Consequently, the mental map between such views cannot be preserved as it is not possible to track and identify changes efficiently.
% Solution 
We provide brushing and linking methods to minimize the cognitive load of identifying nodes in different visual metaphors.
% Minimising temporal aliases
Another limitation of our approach is the fact that specific snapshots can be mistaken for other periods (temporal aliases~\cite{beck2009towards}). 
% Solution 
We aim to overcome such temporal aliases by displaying the period in each snapshot view and the time context bar highlighting the underlying period in the overall temporal context.
% Further investigation is required 
We consider these aesthetic challenges~\cite{beck2009towards} as open possibilities for the development of new methods for the interactive comparison of two or more snapshots at different granularities.
% abstract representations 
%For example, we want to investigate in future work how such mixed visual metaphors impact the overall user experience.
For example, the investigation of how such mixed visual metaphors impact the overall user experiences poses an opportunity for future work.

% - Scalability
% Why
The applied methods during the visual analysis influence the computational and visual scalability of our approach.
% Computational scalability 
For instance, the live computation of displayed graph summaries scales linearly to the number of time steps and the size of the evolving graphs.
% Real-time analysis suffers 
Furthermore, the real-time analysis of snapshots can suffer based on the algorithmic time complexities of applied methods, for example, the Clauset-Newman-Moore community detection algorithm~\cite{clauset2004finding}.
% Possible solution 
A possible solution to these challenges is to investigate how graph embeddings can be utilized to guide an analyst towards temporal changes to speed up the analysis process.
% Visual scalability 
Furthermore, the display space limits the visual scalability and readability of structural properties in a snapshot view since they depend on the number of presented snapshots. 
% Solution
To address this, we limit and automatically abstract the number of depicted snapshot views to provide visually readable representations. 
% Rule of thumb 
The limit for the number of snapshots is adjustable and is as a heuristic limited to six snapshot views for each level.
% Future work 
The visual scalability can also be increased by adapting the visual metaphors based on graph properties, such as, automatically presenting matrix-based visualization for dense graphs.

% Further limitations 
We showed the applicability of the approach through the visual analysis of similar periods in a dynamic hyperlink graph, which required an initial starting point for the similarity search (e.g., the election week). 
% Analyst knowledge
An analyst has to be aware of such states in advance or apply automated analysis methods to identify them, for instance, by using change-point detection~\cite{akoglu2010event} algorithms on the graph embeddings. 
% \ts{this is in fact a good extension; I suggest to iclude change point detection, or even interactive learning of detection, in a next step}
% Usability 
Furthermore, the variety of functionality also affects the usability of the approach since the application prototype can be challenging to use for untrained users.
% Automatic analysis methods 
In general, the usage of user guidance in combination with the potential application of more automatic analysis methods (e.g., outlier detection algorithms), can help to set high-level snapshots in the context of low-level snapshots, drill down the temporal hierarchy, and steer the user towards a useful combination of data and visual transformations to highlight specific trends.
% The hierarchy 
For example, the utilization of sub-queries in the temporal hierarchy can be used to steer an analyst towards fine-grained states with particular graph properties (e.g., motifs).

% Comparison to other approaches 
A limitation of our work is the lack of a formal comparative study to compare Multiscale Snapshots with other visual analytics approaches.
% Approach 
In general, such a comparison remains challenging as our approach allows us to integrate visualization techniques (e.g., van Elzen et al.~\cite{van2016reducing}), which is a simple way to increase the overall task coverage. 
% The evaluation 
Despite the shortage of a comparative study, our quantitative evaluation and the usage scenario highlight key benefits of our approach, such as the multiscale embedding of sequences of graphs to speeds up analytical tasks (e.g., similarity search). 
Graph embeddings come with the sacrifice of information loss compared to methods such as the computation of graph editing distance (GED)~\cite{bunke1983distance}.
% Formal study 
In future work, we aim to overcome shortcomings by integrating new visual metaphors to allow analysts to examine snapshots and their graph embeddings to understand and interpret the quality of the underlying graph embeddings.

%%%%%%%%%%%%%%%%%%%%%%%%%%%%%%%%%%%%%%%%%%%%%%%%%%%%%%%%%%%%%%%%%%%%%%%%%%%
% CONCLUSION
%%%%%%%%%%%%%%%%%%%%%%%%%%%%%%%%%%%%%%%%%%%%%%%%%%%%%%%%%%%%%%%%%%%%%%%%%%%
\section{Conclusion} \label{sec:conclusion}
% Why 
In this paper, we presented Multiscale Snapshots, a visual analytics approach, to provide an overview of a dynamic graph.
% How 
The approach consists of three steps: creating multiscale temporal summaries, applying graph embeddings, and the semi-automatic visual analysis. 
% Main idea 
The combination of the steps enables us to visually explore how temporal and structural properties affect the overall dynamic graph. 
% The approach 
We implemented a prototype and showed in a quantitative evaluation that the approach helps to identify similar temporal states in artificial and real-world dynamic graphs. 
% The use case 
We also show the applicability by a usage scenario analyzing a real-world dataset, demonstrating that patterns in dynamic graphs can be visually analyzed over time.

% Potential future work
The application of Multiscale Snapshots and the underlying multiscale temporal analysis paradigm is not limited to dynamic graphs and can be extended in several ways to work with any temporal data. 
For instance, the Multiscale Snapshots approach can be adjusted to support the user-driven analysis of multivariate time-series data.

%% if specified like this the section will be committed in review mode
\acknowledgments{This work was partly funded by the Deutsche Forschungsgemeinschaft (DFG, German Research Foundation) under Germany's Excellence Strategy - EXC 2117 - 422037984 and the European Union’s Horizon 2020 research and innovation programme under grant agreement No 830892.}

\bibliographystyle{abbrv-doi}

\bibliography{template}

\begin{thebibliography}{10}

\bibitem{abello2004matrix}
J.~Abello and F.~Van~Ham.
\newblock Matrix zoom: A visual interface to semi-external graphs.
\newblock In {\em IEEE symposium on information visualization}, pp. 183--190.
  IEEE, 2004.

\bibitem{aggarwal2015outlier}
C.~C. Aggarwal.
\newblock Outlier analysis.
\newblock In {\em Data mining}, pp. 237--263. Springer, 2015.

\bibitem{ahn2011temporal}
J.-w. Ahn, M.~Taieb-Maimon, A.~Sopan, C.~Plaisant, and B.~Shneiderman.
\newblock Temporal visualization of social network dynamics: Prototypes for
  nation of neighbors.
\newblock In {\em International Conference on Social Computing,
  Behavioral-Cultural Modeling, and Prediction}, pp. 309--316. Springer, 2011.

\bibitem{aigner2008visual}
W.~Aigner, S.~Miksch, W.~M{\"u}ller, H.~Schumann, and C.~Tominski.
\newblock Visual methods for analyzing time-oriented data.
\newblock {\em IEEE transactions on visualization and computer graphics},
  14(1):47--60, 2008.

\bibitem{akoglu2010event}
L.~Akoglu and C.~Faloutsos.
\newblock Event detection in time series of mobile communication graphs.
\newblock In {\em Army science conference}, vol.~1, 2010.

\bibitem{archambault2014temporal}
D.~Archambault, J.~Abello, J.~Kennedy, S.~Kobourov, K.-L. Ma, S.~Miksch,
  C.~Muelder, and A.~C. Telea.
\newblock Temporal multivariate networks.
\newblock In {\em Multivariate Network Visualization}, pp. 151--174. Springer,
  2014.

\bibitem{archambault2008grouseflocks}
D.~Archambault, T.~Munzner, and D.~Auber.
\newblock Grouseflocks: Steerable exploration of graph hierarchy space.
\newblock {\em IEEE transactions on visualization and computer graphics},
  14(4):900--913, 2008.

\bibitem{archambault2011animation}
D.~Archambault, H.~Purchase, and B.~Pinaud.
\newblock Animation, small multiples, and the effect of mental map preservation
  in dynamic graphs.
\newblock {\em IEEE Transactions on Visualization and Computer Graphics},
  17(4):539--552, 2011.

\bibitem{bach2015small}
B.~Bach, N.~Henry-Riche, T.~Dwyer, T.~Madhyastha, J.-D. Fekete, and
  T.~Grabowski.
\newblock Small multipiles: Piling time to explore temporal patterns in dynamic
  networks.
\newblock In {\em Computer Graphics Forum}, vol.~34, pp. 31--40. Wiley Online
  Library, 2015.

\bibitem{bach2014visualizing}
B.~Bach, E.~Pietriga, and J.-D. Fekete.
\newblock Visualizing dynamic networks with matrix cubes.
\newblock In {\em Proceedings of the SIGCHI conference on Human Factors in
  Computing Systems}, pp. 877--886. ACM, 2014.

\bibitem{bach2015time}
B.~Bach, C.~Shi, N.~Heulot, T.~Madhyastha, T.~Grabowski, and P.~Dragicevic.
\newblock Time curves: Folding time to visualize patterns of temporal evolution
  in data.
\newblock {\em IEEE transactions on visualization and computer graphics},
  22:559--568, 2015.

\bibitem{beck2009towards}
F.~Beck, M.~Burch, and S.~Diehl.
\newblock Towards an aesthetic dimensions framework for dynamic graph
  visualisations.
\newblock In {\em 2009 13th International Conference Information
  Visualisation}, pp. 592--597. IEEE, 2009.

\bibitem{beck2017taxonomy}
F.~Beck, M.~Burch, S.~Diehl, and D.~Weiskopf.
\newblock A taxonomy and survey of dynamic graph visualization.
\newblock In {\em Computer Graphics Forum}, vol.~36, pp. 133--159. Wiley Online
  Library, 2017.

\bibitem{bederson1994pad++}
B.~B. Bederson, L.~Stead, and J.~D. Hollan.
\newblock Pad++: Advances in multiscale interfaces.
\newblock In {\em Conference on Human Factors in Computing Systems: Conference
  companion on Human factors in computing systems}, vol.~24, pp. 315--316,
  1994.

\bibitem{behrisch2016matrix}
M.~Behrisch, B.~Bach, N.~Henry~Riche, T.~Schreck, and J.-D. Fekete.
\newblock Matrix reordering methods for table and network visualization.
\newblock In {\em Computer Graphics Forum}, vol.~35, pp. 693--716. Wiley Online
  Library, 2016.

\bibitem{bender2006art}
S.~Bender-deMoll and D.~A. McFarland.
\newblock The art and science of dynamic network visualization.
\newblock {\em Journal of Social Structure}, 7(2):1--38, 2006.

\bibitem{bezerianos2010graphdice}
A.~Bezerianos, F.~Chevalier, P.~Dragicevic, N.~Elmqvist, and J.-D. Fekete.
\newblock Graphdice: A system for exploring multivariate social networks.
\newblock In {\em Computer Graphics Forum}, vol.~29, pp. 863--872. Wiley Online
  Library, 2010.

\bibitem{blumenthal2018exact}
D.~B. Blumenthal and J.~Gamper.
\newblock On the exact computation of the graph edit distance.
\newblock {\em Pattern Recognition Letters}, 2018.

\bibitem{brandes2004analysis}
U.~Brandes and D.~Wagner.
\newblock Analysis and visualization of social networks.
\newblock In {\em Graph drawing software}, pp. 321--340. Springer, 2004.

\bibitem{brehmer2013multi}
M.~Brehmer and T.~Munzner.
\newblock A multi-level typology of abstract visualization tasks.
\newblock {\em IEEE transactions on visualization and computer graphics},
  19(12):2376--2385, 2013.

\bibitem{bunke1983distance}
H.~Bunke.
\newblock What is the distance between graphs.
\newblock {\em Bulletin of the EATCS}, 20:35--39, 1983.

\bibitem{burch2017dynamic}
M.~Burch and T.~Reinhardt.
\newblock Dynamic graph visualization on different temporal granularities.
\newblock In {\em 2017 21st International Conference Information Visualisation
  (IV)}, pp. 230--235. IEEE, 2017.

\bibitem{burch2015flip}
M.~Burch and D.~Weiskopf.
\newblock Flip-book visualization of dynamic graphs.
\newblock {\em Int. J. Software and Informatics}, 9(1):3--21, 2015.

\bibitem{chen2019gl2vec}
H.~Chen and H.~Koga.
\newblock Gl2vec: Graph embedding enriched by line graphs with edge features.
\newblock In {\em International Conference on Neural Information Processing},
  pp. 3--14. Springer, 2019.

\bibitem{chung2019hierarchical}
J.~Chung, S.~Ahn, and Y.~Bengio.
\newblock Hierarchical multiscale recurrent neural networks.
\newblock In {\em 5th International Conference on Learning Representations,
  ICLR 2017}, 2019.

\bibitem{clauset2004finding}
A.~Clauset, M.~E. Newman, and C.~Moore.
\newblock Finding community structure in very large networks.
\newblock {\em Physical review E}, 70(6):066111, 2004.

\bibitem{cui2006measuring}
Q.~Cui, M.~Ward, E.~Rundensteiner, and J.~Yang.
\newblock Measuring data abstraction quality in multiresolution visualizations.
\newblock {\em IEEE Transactions on Visualization and Computer Graphics},
  12(5):709--716, 2006.

\bibitem{cui2014let}
W.~Cui, X.~Wang, S.~Liu, N.~H. Riche, T.~M. Madhyastha, K.~L. Ma, and B.~Guo.
\newblock Let it flow: a static method for exploring dynamic graphs.
\newblock In {\em 2014 IEEE Pacific Visualization Symposium}, pp. 121--128.
  IEEE, 2014.

\bibitem{dal2017wavelet}
A.~Dal~Col, P.~Valdivia, F.~Petronetto, F.~Dias, C.~T. Silva, and L.~G. Nonato.
\newblock Wavelet-based visual analysis of dynamic networks.
\newblock {\em IEEE transactions on visualization and computer graphics},
  24(8):2456--2469, 2017.

\bibitem{diehl2002graphs}
S.~Diehl and C.~G{\"o}rg.
\newblock Graphs, they are changing.
\newblock In {\em International Symposium on Graph Drawing}, pp. 23--31.
  Springer, 2002.

\bibitem{elmqvist2008zame}
N.~Elmqvist, T.-N. Do, H.~Goodell, N.~Henry, and J.-D. Fekete.
\newblock Zame: Interactive large-scale graph visualization.
\newblock In {\em 2008 IEEE Pacific visualization symposium}, pp. 215--222.
  IEEE, 2008.

\bibitem{elmqvist2010hierarchical}
N.~Elmqvist and J.-D. Fekete.
\newblock Hierarchical aggregation for information visualization: Overview,
  techniques, and design guidelines.
\newblock {\em IEEE Transactions on Visualization and Computer Graphics},
  16(3):439--454, 2010.

\bibitem{federico2012visual}
P.~Federico, J.~Pfeffer, W.~Aigner, S.~Miksch, and L.~Zenk.
\newblock Visual analysis of dynamic networks using change centrality.
\newblock In {\em Proceedings of the 2012 International Conference on Advances
  in Social Networks Analysis and Mining (ASONAM 2012)}, pp. 179--183. IEEE
  Computer Society, 2012.

\bibitem{fruchterman1991graph}
T.~M. Fruchterman and E.~M. Reingold.
\newblock Graph drawing by force-directed placement.
\newblock {\em Software: Practice and experience}, 21(11):1129--1164, 1991.

\bibitem{goyal2018dynamicgem}
P.~Goyal, S.~R. Chhetri, N.~Mehrabi, E.~Ferrara, and A.~Canedo.
\newblock Dynamicgem: A library for dynamic graph embedding methods.
\newblock {\em arXiv preprint arXiv:1811.10734}, 2018.

\bibitem{goyal2018graph}
P.~Goyal and E.~Ferrara.
\newblock Graph embedding techniques, applications, and performance: A survey.
\newblock {\em Knowledge-Based Systems}, 151:78--94, 2018.

\bibitem{hadlak2011situ}
S.~Hadlak, H.-J. Schulz, and H.~Schumann.
\newblock In situ exploration of large dynamic networks.
\newblock {\em IEEE Transactions on Visualization and Computer Graphics},
  17(12):2334--2343, 2011.

\bibitem{hadlak2013supporting}
S.~Hadlak, H.~Schumann, C.~H. Cap, and T.~Wollenberg.
\newblock Supporting the visual analysis of dynamic networks by clustering
  associated temporal attributes.
\newblock {\em IEEE Transactions on Visualization and Computer Graphics},
  19(12):2267--2276, 2013.

\bibitem{harrower2003colorbrewer}
M.~Harrower and C.~A. Brewer.
\newblock Colorbrewer. org: an online tool for selecting colour schemes for
  maps.
\newblock {\em The Cartographic Journal}, 40(1):27--37, 2003.

\bibitem{healey2012attention}
C.~Healey and J.~Enns.
\newblock Attention and visual memory in visualization and computer graphics.
\newblock {\em IEEE transactions on visualization and computer graphics},
  18(7):1170--1188, 2012.

\bibitem{henry2006matrixexplorer}
N.~Henry and J.-D. Fekete.
\newblock Matrixexplorer: a dual-representation system to explore social
  networks.
\newblock {\em IEEE transactions on visualization and computer graphics},
  12(5):677--684, 2006.

\bibitem{javed2012stack}
W.~Javed and N.~Elmqvist.
\newblock Stack zooming for multifocus interaction in skewed-aspect visual
  spaces.
\newblock {\em IEEE transactions on visualization and computer graphics},
  19(8):1362--1374, 2012.

\bibitem{kamada1989algorithm}
T.~Kamada, S.~Kawai, et~al.
\newblock An algorithm for drawing general undirected graphs.
\newblock {\em Information processing letters}, 31(1):7--15, 1989.

\bibitem{kerracher2015task}
N.~Kerracher, J.~Kennedy, and K.~Chalmers.
\newblock A task taxonomy for temporal graph visualisation.
\newblock {\em IEEE transactions on visualization and computer graphics},
  21(10):1160--1172, 2015.

\bibitem{kerren2007human}
A.~Kerren, A.~Ebert, and J.~Meyer.
\newblock {\em Human-Centered Visualization Environments: GI-Dagstuhl Research
  Seminar, Dagstuhl Castle, Germany, March 5-8, 2006, Revised Papers}, vol.
  4417.
\newblock Springer, 2007.

\bibitem{kumar2018community}
S.~Kumar, W.~L. Hamilton, J.~Leskovec, and D.~Jurafsky.
\newblock Community interaction and conflict on the web.
\newblock In {\em Proceedings of the 2018 World Wide Web Conference on World
  Wide Web}, pp. 933--943. International World Wide Web Conferences Steering
  Committee, 2018.

\bibitem{liu2018graph}
Y.~Liu, T.~Safavi, A.~Dighe, and D.~Koutra.
\newblock Graph summarization methods and applications: A survey.
\newblock {\em ACM Computing Surveys (CSUR)}, 51(3):1--34, 2018.

\bibitem{malkov2018efficient}
Y.~A. Malkov and D.~A. Yashunin.
\newblock Efficient and robust approximate nearest neighbor search using
  hierarchical navigable small world graphs.
\newblock {\em IEEE transactions on pattern analysis and machine intelligence},
  2018.

\bibitem{moody2004dynamic}
J.~Moody, D.~McFarland, and S.~Bender-deMoll.
\newblock Dynamic network visualization: Methods for meaning with longitudinal
  network movies.
\newblock {\em Retrieved November}, 11:2008, 2004.

\bibitem{narayanan2017graph2vec}
A.~Narayanan, M.~Chandramohan, R.~Venkatesan, L.~Chen, Y.~Liu, and S.~Jaiswal.
\newblock graph2vec: Learning distributed representations of graphs.
\newblock {\em arXiv preprint arXiv:1707.05005}, 2017.

\bibitem{newman2018networks}
M.~Newman.
\newblock {\em Networks}.
\newblock Oxford university press, 2018.

\bibitem{nobre2019state}
C.~Nobre, M.~Meyer, M.~Streit, and A.~Lex.
\newblock The state of the art in visualizing multivariate networks.
\newblock In {\em Computer Graphics Forum}, vol.~38, pp. 807--832. Wiley Online
  Library, 2019.

\bibitem{paranjape2017motifs}
A.~Paranjape, A.~R. Benson, and J.~Leskovec.
\newblock Motifs in temporal networks.
\newblock In {\em Proceedings of the Tenth ACM International Conference on Web
  Search and Data Mining}, pp. 601--610, 2017.

\bibitem{purchase2006important}
H.~C. Purchase, E.~Hoggan, and C.~G{\"o}rg.
\newblock How important is the “mental map”?--an empirical investigation of
  a dynamic graph layout algorithm.
\newblock In {\em International Symposium on Graph Drawing}, pp. 184--195.
  Springer, 2006.

\bibitem{karateclub}
B.~Rozemberczki, O.~Kiss, and R.~Sarkar.
\newblock {Karate Club: An API Oriented Open-source Python Framework for
  Unsupervised Learning on Graphs}.
\newblock In {\em Proceedings of the 29th ACM International Conference on
  Information and Knowledge Management (CIKM '20)}. ACM, 2020.

\bibitem{rufiange2013diffani}
S.~Rufiange and M.~J. McGuffin.
\newblock Diffani: Visualizing dynamic graphs with a hybrid of difference maps
  and animation.
\newblock {\em IEEE Transactions on Visualization and Computer Graphics},
  19(12):2556--2565, 2013.

\bibitem{schulz2013grooming}
H.-J. Schulz and C.~Hurter.
\newblock Grooming the hairball-how to tidy up network visualizations?
\newblock In {\em INFOVIS 2013, IEEE Information Visualization Conference},
  2013.

\bibitem{shneiderman2003eyes}
B.~Shneiderman.
\newblock The eyes have it: A task by data type taxonomy for information
  visualizations.
\newblock In {\em The Craft of Information Visualization}, pp. 364--371.
  Elsevier, 2003.

\bibitem{tominski2008task}
C.~Tominski, G.~Fuchs, and H.~Schumann.
\newblock Task-driven color coding.
\newblock In {\em 2008 12th International Conference Information
  Visualisation}, pp. 373--380. IEEE, 2008.

\bibitem{tversky2002animation}
B.~Tversky, J.~B. Morrison, and M.~Betrancourt.
\newblock Animation: can it facilitate?
\newblock {\em International journal of human-computer studies},
  57(4):247--262, 2002.

\bibitem{van2016reducing}
S.~van~den Elzen, D.~Holten, J.~Blaas, and J.~J. van Wijk.
\newblock Reducing snapshots to points: A visual analytics approach to dynamic
  network exploration.
\newblock {\em IEEE transactions on visualization and computer graphics},
  22(1):1--10, 2016.

\bibitem{verma2017hunt}
S.~Verma and Z.-L. Zhang.
\newblock Hunt for the unique, stable, sparse and fast feature learning on
  graphs.
\newblock In I.~Guyon, U.~V. Luxburg, S.~Bengio, H.~Wallach, R.~Fergus,
  S.~Vishwanathan, and R.~Garnett, eds., {\em Advances in Neural Information
  Processing Systems 30}, pp. 88--98. {Curran Associates, Inc.}, 2017.

\bibitem{wang2019nonuniform}
Y.~Wang, D.~Archambault, H.~Haleem, T.~Moeller, Y.~Wu, and H.~Qu.
\newblock Nonuniform timeslicing of dynamic graphs based on visual complexity.
\newblock In {\em 2019 IEEE Visualization Conference (VIS)}, pp. 1--5. IEEE,
  2019.

\bibitem{wong2009multi}
P.~C. Wong, P.~Mackey, K.~A. Cook, R.~M. Rohrer, H.~Foote, and M.~A. Whiting.
\newblock A multi-level middle-out cross-zooming approach for large graph
  analytics.
\newblock In {\em 2009 IEEE Symposium on Visual Analytics Science and
  Technology}, pp. 147--154. IEEE, 2009.

\bibitem{xu2018exploring}
J.~Xu, Y.~Tao, Y.~Yan, and H.~Lin.
\newblock Exploring evolution of dynamic networks via diachronic node
  embeddings.
\newblock {\em IEEE transactions on visualization and computer graphics}, 2018.

\bibitem{zhang2018network}
D.~Zhang, J.~Yin, X.~Zhu, and C.~Zhang.
\newblock Network representation learning: A survey.
\newblock {\em IEEE transactions on Big Data}, 2018.

\bibitem{zinsmaier2012interactive}
M.~Zinsmaier, U.~Brandes, O.~Deussen, and H.~Strobelt.
\newblock Interactive level-of-detail rendering of large graphs.
\newblock {\em IEEE Transactions on Visualization and Computer Graphics},
  18(12):2486--2495, 2012.

\end{thebibliography}
\end{document}